\documentclass[twocolumn]{article}

\usepackage[margin=0.7in]{geometry}
\usepackage[parfill]{parskip}
\usepackage[utf8]{inputenc}
\usepackage{gensymb}
\usepackage{textcomp}
\usepackage{graphicx}
\usepackage{hyperref}
\usepackage{placeins}
\usepackage{tabularx}
\usepackage{array}
\usepackage{amsmath}
\usepackage{mathptmx}
\usepackage{crop,inputenc,amsmath,amssymb,latexsym,
                float,epsfig,wrapfig,graphicx,stfloats,
                multicol,rotating,multirow,dcolumn,
                boxedminipage,makeidx,soul,url,xspace,times}

\newcommand{\LML}{L_{\text{ML}}}
\newcommand{\LKL}{L_{\text{KL}}}

\newcommand{\Rxz}{R_{xz}}
\newcommand{\Rzx}{R_{zx}}
\newcommand{\Jxz}{J_{xz}}

\newcommand{\Fxz}{F_{xz}}
\newcommand{\Fzx}{F_{zx}}

\newcommand{\KL}{\text{KL}}

\newcommand{\bx}{\mathbf{x}}
\newcommand{\by}{\mathbf{y}}
\newcommand{\bz}{\mathbf{z}}
\newcommand{\bs}{\mathbf{s}}

\newcommand{\bv}{\mathbf{v}}

\newcommand{\bL}{\mathbf{L}}
\newcommand{\bR}{\mathbf{R}}
\newcommand{\ba}{\mathbf{a}}

\newcommand{\br}{\mathbf{r}}
\newcommand{\btheta}{\pmb{\theta}}

\newcommand{\ULJ}{U_{\text{LJ}}}

\newcommand{\Uaug}{U_{\text{aug}}}
\newcommand{\UGaug}{U_G^{\text{aug}}}
\newcommand{\muaug}{\mu^{\text{aug}}_X}

\newcommand{\Gaug}{G^{\text{aug}}_X}

\newcommand{\Deltaaug}{\Delta^{\text{aug}}_X}

\newcommand{\Ucom}{U_{\text{CoM}}}
\newcommand{\kcom}{k_{\text{CoM}}}

\newcommand{\scom}{\mathbf{s}_{\text{CoM}}}

\newcommand{\Clin}{C_{\text{lin}}}
\newcommand{\Cscale}{C_{\text{scale}}}

\newcommand{\rlin}{r_{\text{lin}}}
\newcommand{\rcut}{r_{\text{cut}}}

\newcommand{\nblocks}{n_{\text {blocks }}}
\newcommand{\nlayers}{n_{\text {layers }}}
\newcommand{\nnodes}{n_{\text {nodes }}}

\begin{document}

\title{A Boltzmann generator for the isobaric-isothermal ensemble}

\author{Steyn van Leeuwen,$^{1}$
Alberto P\'erez de Alba Ort\'iz\,$^{1,2}$
and Marjolein Dijkstra\,$^{1}$%
\footnote{Email: m.dijkstra@uu.nl}}

\date{%
\small{$^{1}$Soft Condensed Matter, Debye Institute for Nanomaterials Science, Utrecht University, Princetonplein 1, 3584 CC Utrecht, The Netherlands\\
$^{2}$Computational Soft Matter, van 't Hoff Institute for Molecular Sciences and Informatics Institute, University of Amsterdam, Science Park 904, 1098 XH Amsterdam, The Netherlands.\\[2ex]
\today}
}


\maketitle

\begin{abstract}
Boltzmann generators (BGs) are now recognized as forefront generative models for sampling equilibrium states of many-body systems in the canonical ensemble, as well as for calculating the corresponding Helmholtz free energy. Furthermore, BGs can potentially provide a notable improvement in efficiency compared to conventional  techniques such as  molecular dynamics (MD) and Monte Carlo (MC) methods. By sampling from a clustered latent space, BGs can circumvent free-energy barriers  and overcome the rare-event problem. However, one major limitation of BGs is their inability to sample  across phase transitions between ordered phases. This is due to the fact that  new phases may not be commensurate  with the box dimensions, which remain fixed  in the canonical ensemble. In this work, we present a novel BG model for the isothermal-isobaric ($NPT$) ensemble, which  can successfully overcome this limitation. This unsupervised machine-learning model can sample equilibrium states at various  pressures, as well as pressure-driven phase transitions. We demonstrate that the samples generated by this model are in good agreement with those obtained through MD simulations of two model systems. Additionally, we derive an estimate of the Gibbs free energy using samples generated by the $NPT$ BG.
\end{abstract}

\section{INTRODUCTION}
Boltzmann generators (BGs), \cite{noe_boltzmann_2019} first introduced in 2019, offer a promising new approach for efficiently sampling the  equilibrium states of many-body systems. Traditionally,  sampling the Boltzmann distribution\cite{boltzmann1898vorlesungen} of atomic or colloidal systems relies on a step-wise propagation, either via molecular dynamics (MD)\cite{alder1959studies} or Monte Carlo (MC)\cite{metropolis1953equation} simulations. Although these methods are from the 1950s, across decades they have been immensely boosted by advances in computational power and by enhanced sampling algorithms, \cite{torrie_nonphysical_1977, laio_escaping_2002, henin2022enhanced} which enable to explore and render free-energy landscapes of many-body systems. These accelerated computer experiments are nowadays routinely used to extract valuable insight about complex phenomena occurring in a variety of systems, ranging from molecules, to polymers, to colloids. The ability to sample the Boltzmann distribution is essential for  understanding such many-body systems, because it determines their equilibrium properties. Therefore the simulation community is constantly seeking  for more efficient ways to perform this task in order to tackle increasingly complex systems. Although coordinate transformations were proposed as a solution for  barrier-less sampling, \cite{zhu2002using}  the case-by-case nature of such  transformations impeded  progress. Fortunately, recent advances in machine learning have provided a solution. BGs can  provide one-shot samples from the Boltzmann distribution without the need for time-stepping of MC moves, once a suitable transformation has been machine-learned. As the simulation community starts to invest more in improving BGs, in the same way that MD and MC were boosted  decades ago, we present here an extension of BGs, which to our knowledge is the first BG for the isobaric-isothermal ensemble.

Boltzmann Generators (BGs) were proposed by No\'{e} {\em et al.}  \cite{noe_boltzmann_2019} in 2019  for efficiently sampling the equilibrium states of many-body condensed-matter systems in one shot. BGs leverage normalizing flows (NFs), which are trainable and invertible transformations, to transform a simple prior distribution $\mu_Z$ (e.g. a Gaussian distribution) in latent space $Z$ into the Boltzmann distribution $\mu_X$ in configuration space $X$ through a coordinate transformation $\bx = \Fzx(\bz)$. Once the NF is trained, one can obtain configurations $\bx$ by sampling from the prior distribution ($\bz \sim \mu_Z$) and transforming those samples to configurations $\bx = \Fzx(\bz)$. In this way, BGs bypass the step-wise nature of previous sampling methods and efficiently obtain configurations in one shot. As a result, BGs eliminate the need for an iterative process of updating a single configuration and climbing over free-energy barriers. This innovative approach to sampling provides a powerful and scalable tool for exploring the equilibrium properties of complex systems in various fields.

In recent years, BGs have been improved to handle the large symmetry groups that are associated with many-body systems. \cite{ahmad_free_2022, wirnsberger_normalizing_2021, wirnsberger_targeted_2020, kohler_equivariant_2020, satorras_en_2021,winter_unsupervised_2022} 
However, it is important to note that all these BGs sample many-body systems in the canonical ensemble, i.e. the system is at a constant number of particles $N$,  temperature $T$, and volume $V$. This constant volume constraint can cause problems when sampling across phase transitions, since a crystal may not be commensurate with the fixed box dimensions. 
The isobaric-isothermal ensemble circumvents this problem since the pressure $P$ is held constant instead of the volume, allowing the volume  to fluctuate such that different  crystals may be accommodated. Additionally, this ensemble enables the calculation of the Gibbs free energy as opposed to the Helmholtz free energy in the canonical ensemble.

In this work, we explore how BGs can be extended for sampling many-body systems in the isobaric-isothermal ensemble, i.e. at constant number of particles $N$, pressure $P$, and temperature $T$.  More precisely, we generalize an existing method by Ahmad and Cai\cite{ahmad_free_2022} from the canonical ensemble to the isobaric-isothermal ensemble. Using two case studies, namely a Lennard-Jones (LJ) and a Hemmer-Stell-like (HSL) system, we compare the performance of the new $NPT$ BG to MD simulations,  and observe good agreement between the two methods. Additionally, we also derive an expression to extract the Gibbs free energy from the $NPT$ BG.

\color{black}

\section{METHODS}
\subsection{Isobaric-isothermal ensemble}
\label{sec:NPTtheory}
In the isobaric-isothermal ensemble, we consider configurations with a fixed number of particles $N$, fixed pressure $P$,  and fixed temperature $T$.\cite{frenkel_understanding_2002}  Since $P$ is fixed, the volume $V$ of the system is allowed to fluctuate. Therefore, a configuration in the $NPT$ ensemble is not only described by particle positions $\bx$, but also by the box dimensions $\bL = (L_x,L_y)$ in a two-dimensional, or $\bL=(L_x,L_y,L_z)$ in a three-dimensional system, which are allowed to fluctuate. More precisely, a configuration ($\bL,\bs)$ can be defined in 2D by the box dimensions $\bL = (L_x,L_y)$ and by the scaled positions of all its particles, i.e. $\bs = (\bs_1, ..., \bs_N)$, where $\bs_i = (x_i/L_x,y_i/L_y)$ denotes the coordinates of particle $i$ scaled by the box dimensions. The probability of a configuration $(\bL,\bs)$ is given by 
\begin{equation}
    \mu_X(\bL,\bs) \propto V^{N} e^{-\beta P V}  e^{-\beta U(\bs, \bL)},
\end{equation}
 which we refer to as the isobaric-isothermal probability distribution, and 
where $V$ denotes the volume of the system, i.e. $V=L_x L_y$ in 2D. Furthermore, $U(\bx)$ is the potential energy of configuration $\bx$ and $\beta = 1/k_B T$ with $k_B$ the Boltzmann constant. 

The isobaric-isothermal partition function $\Delta$ is given by
\begin{equation}
\begin{split}
    \Delta(N,P,T)
    &=\frac{C}{\Lambda^{D N} N !} \int \,d V\ V^N e^{-\beta P V} \int \,d \bs^N\ e^{-\beta U(\bs,\bL)}, \\
    &= \frac{C}{\Lambda^{D N} N !} \Delta_X,
\end{split}
\label{eq:partfuncNPT}
\end{equation}
where $C$ is some constant with units of inverse volume, usually set to $C = \beta P$, and $\Delta_X$
is the configurational partition function. The Gibbs free energy $G$ can be expressed in terms of $\Delta$ as 
\begin{equation}
\begin{split}
    \beta G 
    = - \log(\Delta)
    = - \log\left(\frac{\beta P}{\Lambda^{D N} N !} \right) + \beta G_X, 
    \end{split}
\label{eq:Gtot}
\end{equation}
where $\beta G_X = - \log(\Delta_X)$ is the configurational contribution to the Gibbs free energy. 
The difficulty in computing $\Delta$ and $G$ lies in computing $\Delta_X$ and $G_X$, i.e. the  configurational contributions. In this work, we focus on estimating $\Delta_X$ and $G_X$ by generating samples according to the isobaric-isothermal probability distribution
\begin{equation}
\begin{split}
    \mu_X(\bL,\bs) 
    &= \dfrac{1}{\Delta_X}  e^{-\beta U(\bL,\bs)-\beta P V + N \log(V)}.
    \label{mu_X_complete}
\end{split}
\end{equation}
Furthermore, we are interested in computing ensemble averages of observables over the isobaric-isothermal probability distribution. The average of an observable $O$ over $\mu_X$ is given by
\begin{equation}
\begin{aligned}
&\left<O(\bL,\bs)\right>_{(\bL,\bs) \sim \mu_X} =\int \,d\bL \,d\bs\ \mu_X(\bL, \bs) O(\bL,\bs).
\end{aligned}
\end{equation}

\subsection{Boltzmann generators}
\label{sec:bgs}
\begin{figure}[h!]
    \centering
    \includegraphics[width=0.8\linewidth]{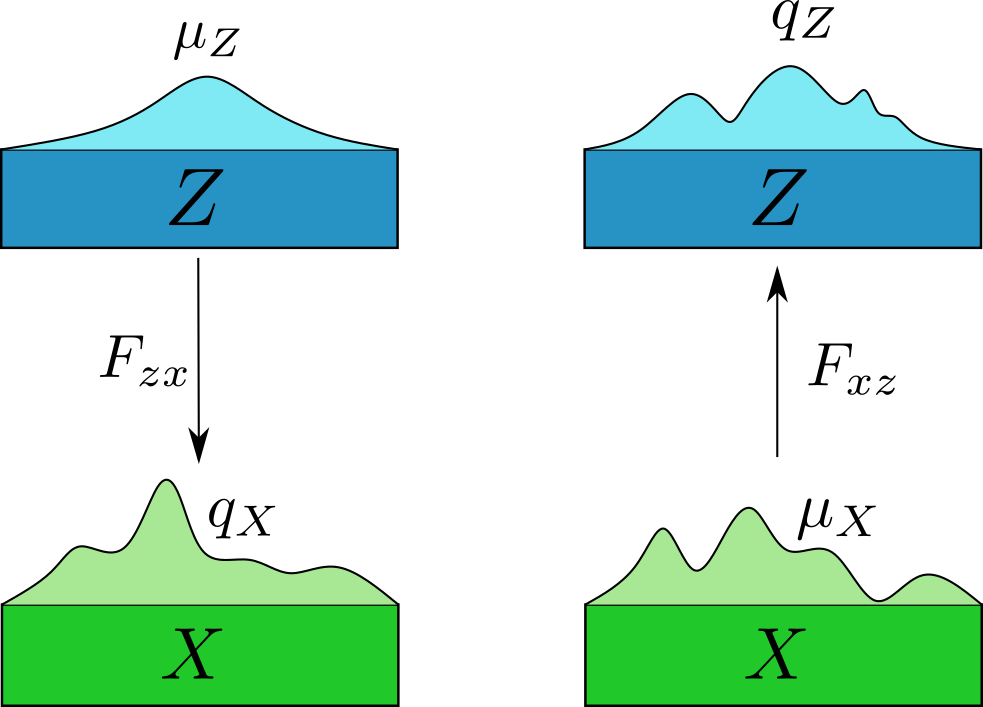}
    \caption{The main idea behind a NF is the transformation of probability distributions. Samples in latent space $Z$ are generated by sampling from a simple distribution $\mu_Z(\bz)$ (e.g. a Gaussian distribution). Then a trainable and invertible transformation $F_{zx}$ maps the samples in $Z$ to samples in configuration space $X$. Samples in $X$ follow a distribution $q_X(\bx)$ which is similar to the target distribution $\mu_X$ (e.g. a Boltzmann distribution). Furthermore, samples in $X$ (obtained from MD or MC simulation) can also be mapped to $Z$ by $F_{xz}$. This results in a distribution $q_Z$ which is similar to $\mu_Z$. Both directions can be used to train the invertible transformation $\Fxz =(\Fzx)^{-1}$.}
    \label{fig:BGarch}
\end{figure}

The primary goal of a BG is to sample configurations of a particle system without the step-wise propagation used in MD and MC methods. The aim is to directly draw samples of particle configurations, denoted by $\bx$, from a configuration space $X$ that follows the Boltzmann distribution $\mu_X(\bx)$. However, this is a challenging task, as the  distribution is not known a priori. Despite the unknown distribution, it is possible to  calculate the likelihood of a specific sample $\bx$, and use that to our advantage. Specifically, one can sample from a simple, known distribution in a latent space $Z$ and then learn to transform a sample $\bf{z}$ into a sample $\bx$, which follows the Boltzmann distribution. This transformation, known as an NF, denoted as $\bx = \Fzx(\bz)$, can be machine-learned. 

Let us now describe in more detail how NFs generate samples. First, samples are drawn from a simple prior probability distribution $\mu_Z(\bz)$, which could be a Gaussian distribution,  in latent space $Z$, see Fig. \ref{fig:BGarch}. These samples in $Z$ are then transformed into samples in configuration space $X$ through a machine-learned invertible and differentiable transformation $\bx = \Fzx(\bz)$. If $F_{zx}$ has been learned correctly, the resulting samples in $X$ follow a distribution $q_X(\bx)$ that closely resembles  the Boltzmann distribution $\mu_X(\bx)$.
Moreover, the transformation $\Fzx$ can also be inverted as $\Fxz =(\Fzx)^{-1}$, allowing training in both directions. We note that the subscripts ${xz}$ or ${zx}$ indicate the direction of the transformation, i.e. $X$-to-$Z$ or $Z$-to-$X$, respectively
\color{black}

BGs transform  samples in latent space to configuration space  by means of an NF. An NF is a series of learnable, invertible and differentiable transformations
\begin{equation}
\Fzx = f_n \circ ... \circ f_2 \circ f_1.
\label{eq:F}
\end{equation}
The transformation $\Fzx$ generates the new distribution by translating and locally compressing or stretching space. One of the advantages of BGs is that the probability $q_X(\bx)$ of a generated sample $\bx$ can be directly evaluated. More precisely, $q_X(\bx)$ can be related to the prior probability distribution $\mu_Z(\bz)$ through a  change of variables
\begin{equation}
    q_X(\bx) = \mu_Z(F_{xz}(\bx)) \Rxz(\bx),
    \label{eq:changevariables}
\end{equation}
where  $\Rxz(\bx) = |\det (\Jxz(\bx))| = |\det \nabla_{\bx} \Fxz(\bx,\btheta)|$. Note that $q_X(\bx)$ depends on the likelihood of the prior distribution $\mu_Z(F_{xz}(\bx))$, but also on how much the space is locally compressed or stretched, which is quantified by $\Rxz(\bx)$. Furthermore, note that it is necessary for $\Fxz$ to fulfill certain conditions: $\Fxz$ must be invertible, because this ensures $\bx$ to be mapped to a unique $\bz = \Fxz(\bx)$, and $\Fxz$ must be differentiable so that $\Rxz$ is well-defined. Finally, we remark that the advantage of such an exact-likelihood generative model is that $q_X$ can be used to reweight samples.
 
Additionally, Fig. \ref{fig:BGarch} shows that $\Fxz =(\Fzx)^{-1}$ can also be used to map configurations from $X$ to $Z$. This allows us to train the NF in two ways.\cite{noe_boltzmann_2019} The first way is by sampling a batch $B$ from $\mu_Z$, mapping $B$ to $X$ via $\Fzx$ and improving $\Fzx$ by minimizing a loss function that is based on the difference between $q_X(\bx)$ and $\mu_X(\bx)$ for all $\bx \in \Fzx(B)$. The second way is by using an MC simulation to sample a batch $B$ from $\mu_X$, mapping $B$ to $Z$ via $\Fxz$ and updating $\Fxz$ through  a loss function  which minimizes the difference between $q_Z(\bz)$ and $\mu_Z(\bz)$ for all $\bz \in \Fxz(B)$. More details on the loss functions used for training the NFs are presented in Section \ref{sec:loss_npt}.

\begin{figure}[h!]
     \centering
     \includegraphics[width=0.8\linewidth]{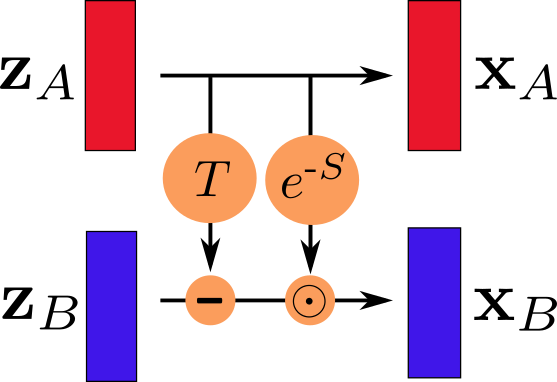}
     \caption{The affine coupling layers in the RealNVP blocks. The input variable $\bz$ is split into two channels: $\bz_A$ and $\bz_B$. The first channel is simply copied, while the second channel is transformed using an affine transformation that uses the first channel as input.}
     \label{fig:nvp}
 \end{figure}

Ideally, NFs should be easy to compute and  invert,  and the determinant of their Jacobian should be easy to calculate. There are many implementations that meet these requirements,   \cite{papamakarios_normalizing_2021, kobyzev_normalizing_2021} such as coupling flows,  \cite{ahmad_free_2022, noe_boltzmann_2019} auto-regressive flows, residual flows,  \cite{wirnsberger_normalizing_2021, wirnsberger_targeted_2020} and infinitesimal flows. \cite{satorras_en_2021, kohler_equivariant_2020} In this work, we use the real-valued non-volume preserving (RealNVP) coupling layers \cite{dinh_density_2017} to implement $f_i$ in Eq. \ref{eq:F}. In a RealNVP coupling layer, the input variable $\bz$ is split into two channels: $\bz_A = \bz_{1:d}$, $\bz_B = \bz_{d+1:D}$ (See Fig. \ref{fig:nvp}). Here, $D$ is the dimensionality of the system and $1 \leq d < D$. The coupling layer does not transform $\bz_A$ and just copies it to obtain the output $\bx_A$. However, $\bz_B$ is transformed using an affine transformation that uses $\bz_A$ as input
\begin{equation}
    f_{zx}(\bz_A,\bz_B) = 
    \begin{cases}
        \bx_A &= \bz_A \\
        \bx_B &= \exp(-S(\bz_A)) \odot \left(\bz_B - T(\bz_A)\right),
    \end{cases}
    \label{eq:affinecoup}
\end{equation}
where $S: \mathbb{R}^{d} \rightarrow \mathbb{R}^{D-d}$ and $T: \mathbb{R}^{d} \rightarrow \mathbb{R}^{D-d}$ are machine-learnable functions (usually deep neural networks), $\odot$ denotes element-wise multiplication and $\exp$ is applied element-wise to $-S(\bz_A)$. To transform all coordinates, the roles of $\bz_B$ and $\bz_A$ are switched in the next affine coupling layer so that $\bz_A$ is transformed and $\bz_B$ is copied. This combination of two affine coupling layers is referred to as a RealNVP block. 

This scheme has all the properties we desire. It is easy to compute since it is just a linear transformation. Furthermore, it is easy to invert because $\bz_A$ is not transformed, so that $S(\bx_A)=S(\bz_A)$ and $T(\bx_A)=T(\bz_A)$ can be readily computed and therefore $S$ and $T$ do not need to be inverted. Therefore the inverse transformation is given by
\begin{equation}
    f_{xz}(\bx_A, \bx_B) = 
    \begin{cases}
        & \bz_A = \bx_A,\\
        & \bz_B = \exp(S(\bx_A)) \odot \bx_B + T(\bx_A), 
    \end{cases}
    \label{eq:affinecoupxz}
\end{equation}
and finally, the Jacobian of $f_{zx}$ and $f_{xz}$ can be easily computed. For $f_{xz}$, the Jacobian is given by
\begin{equation}
\begin{split}
    \nabla_\bx f_{xz}(\bx_A,\bx_B) &=
     \begin{pmatrix}
        \partial_{\bx_A} \bz_A & \partial_{\bx_B} \bz_A \\
        \partial_{\bx_A} \bz_B & \partial_{\bx_B} \bz_B
    \end{pmatrix}, \\
    &=
    \begin{pmatrix}
        \mathbb{I} & 0\\
        \partial_{\bx_A} \bz_B & \text{diag}(\exp{S(\bx_A)}) 
    \end{pmatrix},
    \label{eq:jaco}
\end{split}
\end{equation}
and its Jacobian can be written as $|\det \nabla_\bx f_{xz}| = \prod_{i=1}^{D-d} \exp{S_i(\bx_A)}$. Furthermore, the logarithm of the determinant is given by
\begin{equation}
    \log |\det \nabla_\bx f_{xz}| = \sum_{i=1}^{D-d} S_i(\bx_A),
\end{equation}
which can be computed easily.
To summarize,  BGs use NFs to generate samples according to the Boltzmann distribution. NFs are differentiable bijections that transform a simple distribution $\mu_Z$, e.g. a Gaussian, into a complex distribution $q_X$ which approximates some desired distribution $\mu_X$, e.g. the Boltzmann distribution. These NFs can be used to generate samples according to $\mu_X$ by sampling from $\mu_Z$ and mapping these samples to $X$ via $\Fzx$. Furthermore, we can compute the exact likelihood $q_X$ for all samples generated by the NF. Finally, the NF can be trained in two ways: by improving $\Fzx$ and by improving $\Fxz$.

\subsection{Isobaric-isothermal Boltzmann generator\label{sec:npt_bg}}
\begin{figure}[b!]
\raggedright \large \hspace{0.1cm}(a)\\
\vspace{0.1cm}
\includegraphics[width=0.6\linewidth]{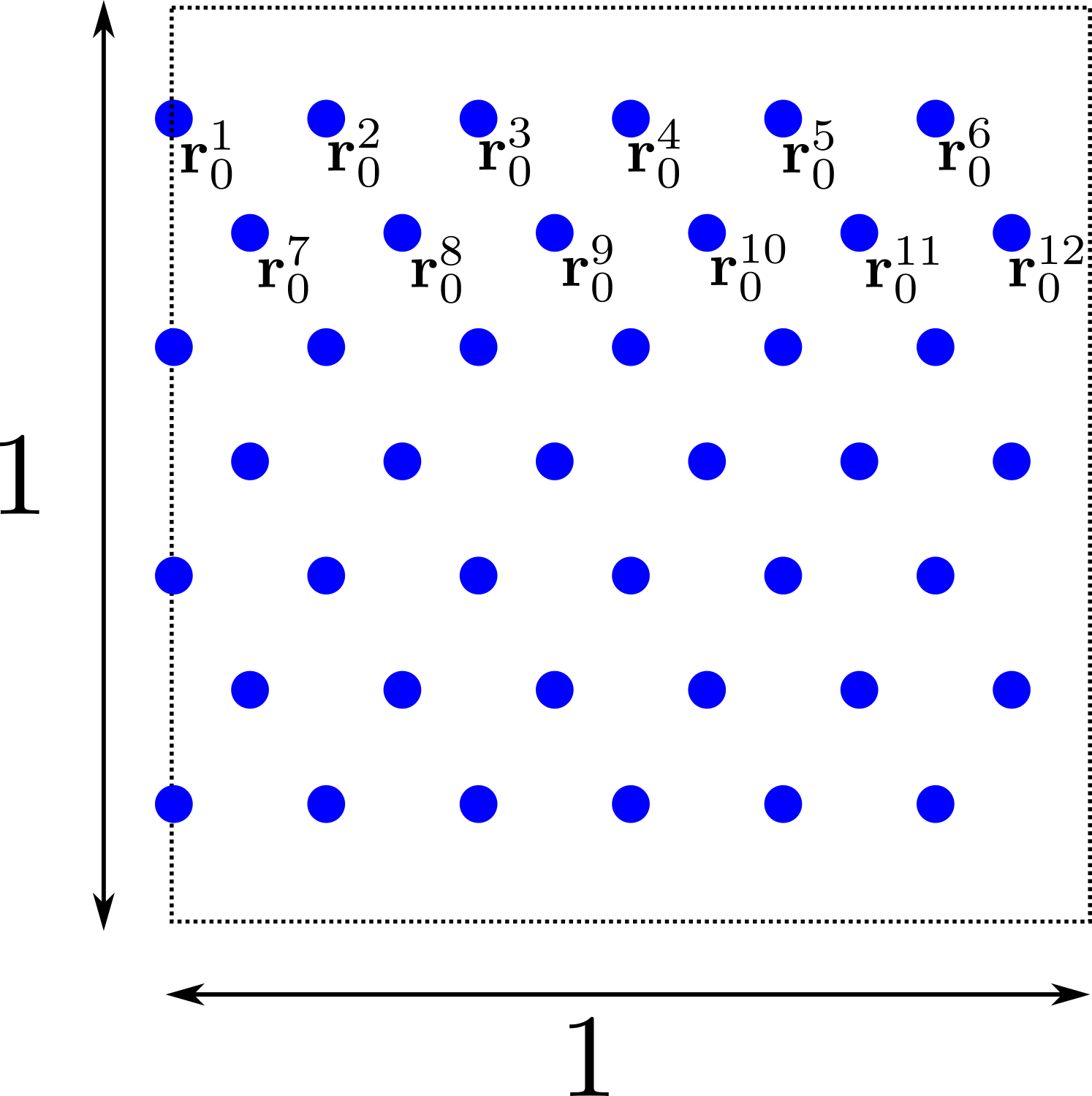}\\
\hspace{0.1cm}(b)\\
\includegraphics[width=0.6\linewidth]{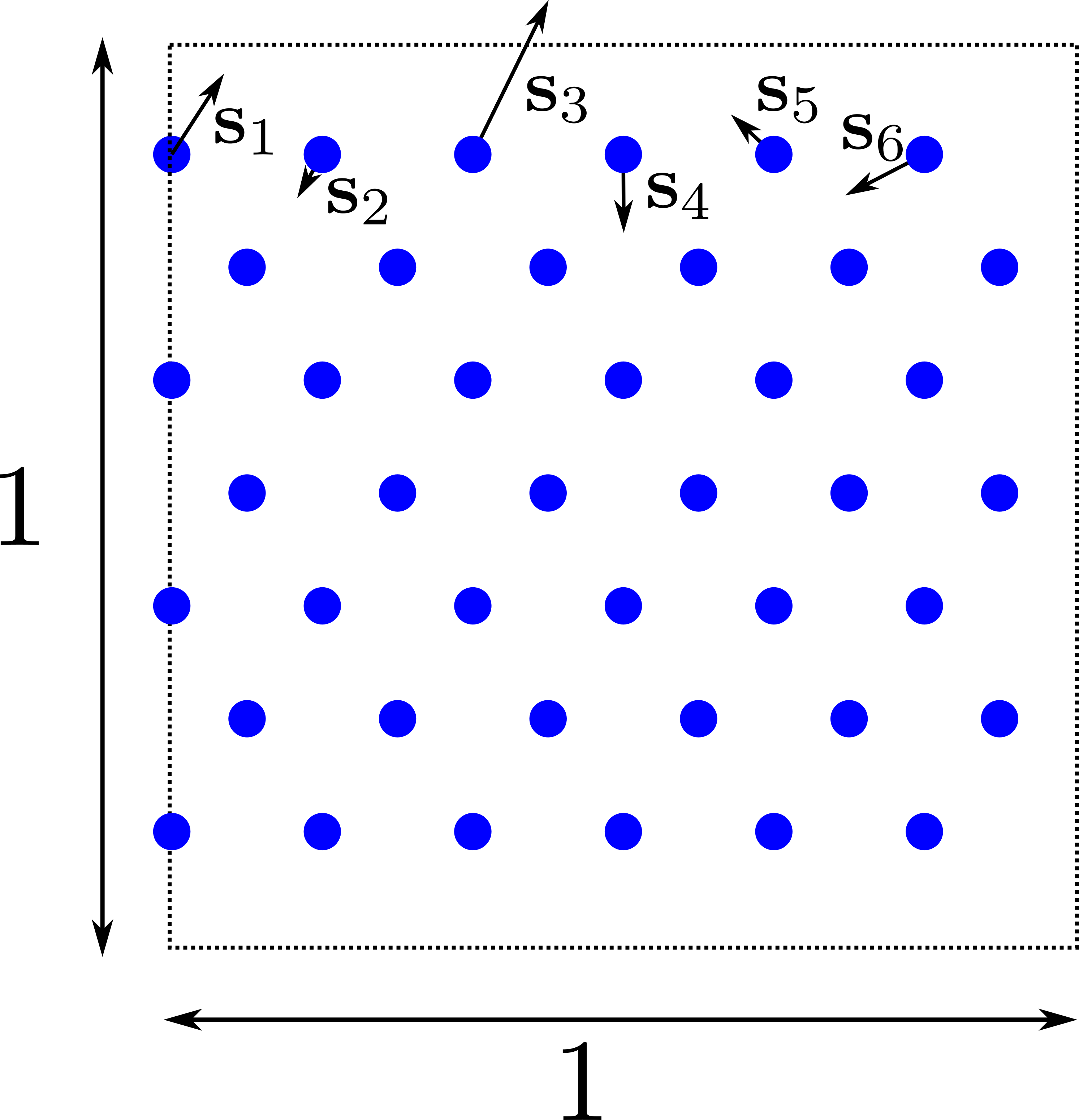}\\
\hspace{0.1cm}(c)\\
\includegraphics[width=0.9\linewidth]{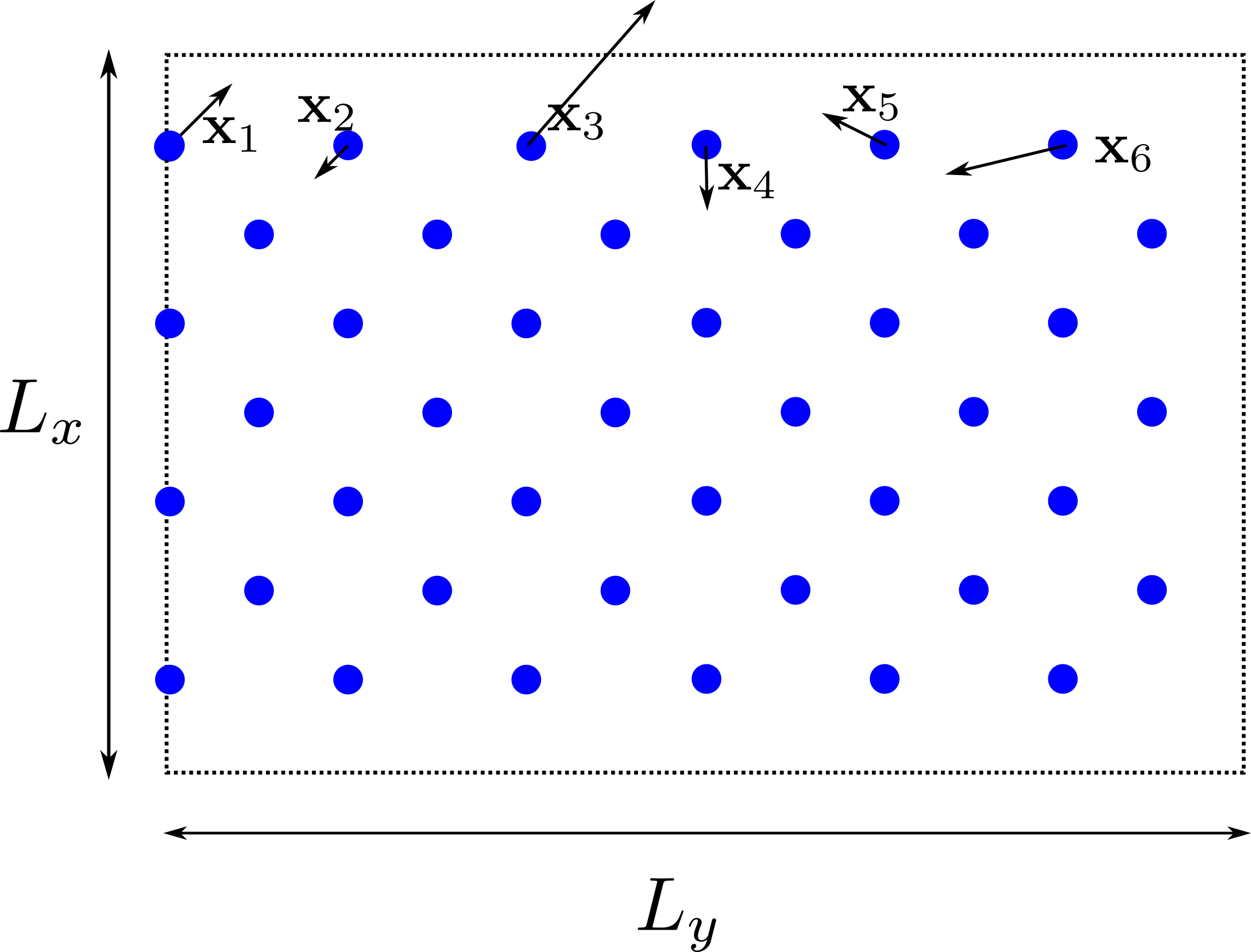}        
\caption{The $NPT$ BG generates scaled deviations $\bs=(\bs_1,...,\bs_N)$ with respect to a scaled reference lattice $\br_0=(\br_0^1,...,\br_0^N)$. (a) The reference lattice $\br_0$. Note that the lattice spacing of $\br_0$ is smaller in the $x$-direction than in the $y$-direction. This is because there are more particle layers in the $x$-direction. (b) The deviations $\bs$ with respect to $\br_0$. Note that $\bs_3$ places particle $3$ outside the box. The particle can be mapped back into the box by applying $ \mathrm{mod} \ 1$ to $\br_0^3+\bs_3$. (c) absolute particle coordinates and box sizes, obtained by scaling by $\bL=(L_x, L_y)$. 
}
\label{fig:npt_deviations}
\end{figure}

In this section, we introduce the isothermal-isobaric Boltzmann generator ($NPT$ BG) for a system in 2D. Inspired by the work of Ahmad and Cai\cite{ahmad_free_2022} we generate displacements from a reference lattice, rather than absolute coordinates. The $NPT$ BG generates configurations $(\bL, \bs)$ according to the $NPT$ distribution with isotropic box fluctuations. Here, $\bL=(L_x,L_y)$ is the box dimension in the $x$-direction and $y$-direction, and $\bs=(\bs_1,...,\bs_N)$ are scaled deviations with respect to a scaled lattice $\br_0=(\br_0^1,...,\br_0^N)$ (see Fig. \ref{fig:npt_deviations}).  The absolute coordinates of the system $\bR=(\bR_1,...,\bR_N)=(R_{1x}. R_{1y},...,R_{Nx},R_{Ny})$ are given by
\begin{equation}
    \begin{split}
        R_{i\alpha}
        &= L_{\alpha} ((s_{i\alpha} + r_0^{i\alpha})\ \mathrm{mod} \ 1),
    \end{split}
\label{eq:NPT_abspos}
\end{equation}
where $i$ is the particle number, $\alpha=x,y$ and $\mathrm{mod} \ 1$ applies the periodic boundary conditions  to the scaled coordinate $s_{i\alpha} + r_0^{i\alpha}$. 

 Any generated vector $\bv \in X$ first lists $\bL$ and then all coordinates  $\bs$, such that $\bv = (\bL,\bs_1,...,\bs_N)$. Thus, Eq. \ref{eq:changevariables} becomes
\begin{equation}
    \begin{split}
        q_X(\bL,\bs)
        =\mu_Z\left(F_{xz}(\bL,\bs)\right) R_{xz}(\bL,\bs),
    \end{split}
    \label{eq:qx_ml_loss}
\end{equation}
where $R_{xz}(\bL,\bs) = |\det(\nabla_{(\bL,\bs)} \Fxz(\bL,\bs))|$ and $\mu_{Z}(\bz)= \exp(-|\bz|^2/ 2\sigma^{2})/\Delta_Z$. 
Note that, since the box fluctuations are isotropic, box proportions are fixed (i.e. $L_y = \lambda L_x$ for some constant $\lambda$). Therefore, it is sufficient to generate only $L_x$ and not the entire vector $\bL=(L_x, L_y)$. Furthermore, the area $V$ can be computed via $V = \lambda L_x^2$ in 2D. In practice, within the NF, we add one more dimension to the latent space $Z$ and configuration space $X$, so that $\dim(Z)=\dim(X)=ND+1$. 

It should be noted that the $NPT$ BG can generate unbounded displacements, as well as configurations with $L_x < 0$. To prevent this, we propose an NF $\Fzx:Z\rightarrow X$ with three transformations, given by
\begin{equation}
    \Fzx
    = F_{\text{lin}} \circ F_{\text{scale}} \circ F_{\text{RealNVP}} ,
\end{equation}
where $F_{\text{RealNVP}}:\mathbb{R}^{ND+1} \rightarrow \mathbb{R}^{ND+1}$ is a standard NF with the RealNVP architecture, $F_{\text{scale}}: \mathbb{R}^{ND+1} \rightarrow \mathbb{R}^{ND+1}$ is a scaling layer and $F_{\text{lin}}$ is a linear translation. More precisely, $F_{\text{scale}}$ is given by 
\begin{equation}
    F_{\text{scale}}(\by) = \Cscale \by,
\end{equation}
with $\Cscale$ a machine learnable parameter, and $F_{\text{lin}}$ reads 
\begin{equation}
    F_{\text{lin}}(\by')
    = \by' + (\Clin,0,...,0),
\end{equation}
where $\Clin$ is also a machine learnable parameter and $\by$ and $\by'$ denote the intermediate coordinate transformations from $\bz$ to $\bx$.

$F_{\text{scale}}$ must be initialized such that the displacements are not larger than half the lattice spacing, i.e. $|\bs_i|<a_{\text{scaled}}/2$ for all $i$. To this end, we first set $\Cscale=1$ and $\Clin=0$ and generate a batch $B$ of configurations $\bv = (\bL,\bs)$. Then we compute the maximum over all coordinates $v_i$ of $\bv$ over all $\bv \in B$ (i.e., $v_{\text{max}} = \max_{\bv \in B} \max_{ 1\leq i\leq ND+1 } |v_i|$). We then set $\Cscale = 1/(\gamma*v_{max})$, where $\gamma$ is a constant that should be chosen such that $1/\gamma < 1/2\sqrt{N}$. This ensures that initial deviations will have a maximum around $1/\gamma$ and that deviations larger than $1/2\sqrt{N}$ are very unlikely. Since $a_{\text{scaled}} \sim 1/\sqrt{N} $, this condition also ensures that $|\bs_i|<a_{\text{scaled}}/2$ for all $i$.

On the other hand, $F_{\text{lin}}$ must be initialised such that the BG does not generate configurations with $L_x<0$. We perform MD simulations and initialise $\Clin$ to approximate the $L_x$ that was observed in these simulations. Therefore, $F_{\text{lin}}$ ensures that the initial distribution of $L_x$ is centered around $\Clin$. Furthermore, $F_{\text{scale}}$ ensures that the maximum deviation from $\Clin$ is $1/\gamma < 1/\sqrt{N}$. Therefore, the minimum $L_x$ that can be generated is around $\Clin - 1/\gamma$, which is much larger than $0$ for the systems we consider. 

To ensure that we sample only one permutation and remove translational invariance, we define an augmented system with a center-of-mass restraint
\begin{equation}
    \Uaug(\bL, \bs) = U(\bL, \bs) + \dfrac{1}{2} \kcom |\scom|^2,
    \label{eq:Uaug}
\end{equation}
where $\bs$ are deviations with respect to $\br_0$, $\kcom$ is the strength of the harmonic center-of-mass potential and $\scom$ is the displacement vector of the center of mass, which is given by
\begin{equation}
    \begin{split}
        \scom = \dfrac{1}{N} \sum_{i=1}^N \bs_i.
    \end{split}
\end{equation}
Note that it is not necessary to apply periodic boundary conditions.

To select an appropriate value for $\kcom$, we examine the average energy resulting from the harmonic potential in $2D$
\begin{equation}
    U_{\text{avg}}
    = N \dfrac{D}{2} k_B T = N k_B T,
\end{equation}
and require that the energy of the harmonic potential at the largest possible deviation is much larger than the average energy of the system
\begin{equation}
    \begin{split}
        \Ucom(1/2\sqrt{N})= \dfrac{1}{2} \kcom (1/2\sqrt{N})^2 &\gg U_{\text{avg}}=N k_b T,\\
        \kcom &\gg 8 N^2 k_B T.
    \end{split}
\label{eq:kcom_min}
\end{equation}
To summarize, our $NPT$ BG generates scaled deviations $\bs$ w.r.t. a fixed reference lattice $\br_0$. The initial deviations can be kept small by using a scaling layer in the NF. This ensures that only one permutation is sampled. Furthermore, the $NPT$ BG generates the box size $\bL$. Absolute particle coordinates can be obtained by multiplying the scaled particle coordinates ($\br_0 + \bs$) by $\bL=(L_x,L_y)$.

Our implementation of BGs is partially based on a repository by Hsu and Fobe,  \cite{hsu_boltzmann_2022} who implemented a double-well potential that we extended to other systems. Additionally, we used utilities from the seminal work on BGs by No\'{e} {\em et al.} \cite{noe_boltzmann_2019} and their repository on BGs and NFs. \cite{kramer_bgflow_2022}
\color{black}

\subsection{Loss functions in the isothermal-isobaric ensemble}
\label{sec:loss_npt}
In this section, we derive the Maximum likelihood (ML) and the Kullback-Leibler (KL) loss for the $NPT$ ensemble. 
The training of our invertible transformations, $\Fzx$ and $\Fxz$, proceeds by  minimizing  specific loss functions. In particular, when learning $\Fxz$, we use the maximum likelihood (ML) loss, which  measures the difference between the approximate  probability distribution $q_Z$ and the ground-truth distribution $\mu_Z$. Conversely, when learning $\Fzx$, we use the Kullback-Leibler (KL) loss, which measures the difference between the approximate probability distribution $q_X$ and the ground-truth distribution $\mu_X$. While we discuss the ML loss for completeness, in this work we focus solely on training based on the KL loss   to use the BG in an unsupervised
way.

\subsubsection{Maximum likelihood loss}
\label{eq:MLlossNPT}
To derive the ML loss function, we take the logarithm of Eq. \ref{eq:qx_ml_loss} and write 
\begin{equation}
    \begin{split}
        \log q_X(\bL,\bs) 
        &= \log(\mu_{Z}\left(F_{xz}(\bL,\bs)\right)) + \log R_{xz}(\bL,\bs), \\
        &=-\log Z_Z - \frac{|\Fxz(\bL,\bs)|^{2}}{2\sigma^{2}}+\log R_{x z}(\bL,\bs),
    \end{split}
\end{equation}
where $R_{xz}(\bL,\bs) = |\det(\nabla_{(\bL,\bs)} \Fxz(\bL,\bs))|$ and $\mu_{Z}(\bz)= \exp(-|\bz|^2/ 2\sigma^{2})/Z_Z$, with $\sigma$ the standard deviation of the Gaussian prior  and $Z_Z$ the normalization factor of the Gaussian prior. We can then write the KL-divergence between $\mu_X$ and $q_X$ as
\begin{eqnarray}
      \hspace{0cm}  \KL \left(\mu_{X} \mid q_{X} \right) &
         \nonumber \\
       & \hspace{-3.0cm} = \int \,d\bL \,d\bs\ \mu_X(\bL,\bs)\left[\log \mu_X(\bL,\bs)-\log q_{X}(\bL,\bs)\right],  \nonumber \\
        & \hspace{-3.5cm} = \mathbb{E}_{(\bL,\bs) \sim \mu_{X}} \left[\frac{\left|F_{xz}(\bL,\bs)\right|^{2}}{2\sigma^{2}}-\log R_{x z}(\bL,\bs)\right] + \text {c},
\end{eqnarray}
where, in the last line, we write $\int \,d\bL \,d\bs\ \mu_X(\bL,\bs) \left[\log(\mu_X(\bL,\bs)) + \log(Z_Z) \right]$ as a constant term $c$ because they do not depend on machine learnable parameters. The ML loss is then defined as
\begin{equation}
    \LML 
    = \mathbb{E}_{(\bL,\bs) \sim \mu_{X}} \left[\frac{\left|F_{xz}(\bL,\bs, \btheta)\right|^{2}}{2\sigma^{2}}-\log R_{x z}(\bL,\bs, \btheta)\right],
    \label{eq:LML_NPT1}
\end{equation}
where $\btheta$ are the machine learnable parameters in $\Fxz$. We can use samples from MC simulations or MD simulations to approximate $\mu_X$.

\subsubsection{Kullback-Leibler loss in the $NPT$ ensemble}
\label{sec:KLlossNPT}
We now derive the KL loss. We start by noting that $q_Z(\bz) = \mu_X(\Fzx(\bz)) \Rzx(\bz)$ and substituting $\mu_X$ from Eq. \ref{mu_X_complete}, such that
\begin{equation}
\begin{aligned}
    \log q_Z(\bz) 
    =& \log(\mu_X(\Fzx(\bz))) + \log(\Rzx(\bz)), \\
    =& -\log(\Delta_X) -\beta U(\bL,\bs)-\beta PV \\
    & + N\log(V) + \log(\Rzx(\bz)). 
\end{aligned}
\end{equation}
We write the KL-divergence between $p=\mu_Z$ and $q=q_Z$ as
\begin{equation}
    \begin{split}
        \KL(\mu_Z|q_Z) \\
      \hspace{-2cm}  = & \int\,d\bz\ \mu_Z(\bz) \left[\log(\mu_Z(\bz)) - \log(q_Z(\bz))\right], \\
        = & \int \,d\bz\ \mu_Z(\bz) [\log(\mu_Z(\bz)) + \log(\Delta_X) \\
         & + \beta U(\bL,\bs)+\beta P V - N\log(V) - \log(\Rzx(\bz))   ], \\
        = & \mathbb{E}_{\bz \sim \mu_Z} [\beta U(\bL,\bs)+\beta P V - N\log(V) \\ 
         & - \log(\Rzx(\bz)) ] + c ,
    \end{split}
\end{equation}
where we have written $\int \,d\bz\ \mu_Z(\bz) \left[\log(\mu_Z(\bz)) + \log(\Delta_X) \right]$  as a constant term $c$ since it does not depend on $\btheta$. Note that $V$ depends on $\bL$. The KL loss can then be defined as
\begin{eqnarray}
    \LKL =\nonumber \\
    && \hspace{-1.8cm} \mathbb{E}_{\bz \sim \mu_Z} \left[\beta U(\bL,\bs)+\beta P V - N\log(V) - \log(\Rzx(\bz, \btheta)) \right].
\label{eq:LK}
\end{eqnarray}

\subsubsection{Gibbs free energy}
\label{sec:G_reweighting_nvpt}
When a physical process takes place under constant temperature and pressure, it can be described in terms of the Gibbs free energy. The Gibbs free energy is a crucial thermodynamic quantity for studying phase transitions in many-body systems since the difference between the Gibbs free energy of two states and the Gibbs free-energy barrier determine whether a transition can occur spontaneously. In this section, we derive the configurational contribution
to the Gibbs free energy $G_X$ using samples from the probability distribution $q_X$ generated by the $NPT$ BG. This derivation can be divided into three steps. First,
based on the augmented system defined in Eq \ref{eq:Uaug}, 
we express the free energy $\Gaug$ of the distribution $\muaug$ by using free-energy perturbation,\cite{zwanzig1954high} a technique in which the thermodynamic properties of one system can be calculated based on a slightly different system and on the difference between the interparticle potentials of the two systems. Second, we calculate the Gibbs free-energy difference between the augmented system, i.e. with the center-of-mass restraint defined in Eq. \ref{eq:Uaug}, and the non-augmented system, i.e. without the center-of-mass restraint. Finally, we use the previous two steps to compute the free energy $G_X$ of $\mu_X$.

Let us first define the Gibbs free energy $\Gaug$ of the augmented system whose energy is given by Eq. \ref{eq:Uaug}. This Gibbs free energy can be expressed in terms of the partition function $\Deltaaug$ as 
\begin{equation}
    \beta \Gaug = - \log(\Deltaaug),
\label{eq:Greweigh1}
\end{equation}
and $\Deltaaug$ can be estimated using samples from $q_X$ as follows
\begin{equation}
\begin{aligned}
        \Deltaaug \\
        & \hspace{-0.6cm} = \int \,d \bL\ \int \,d\bs\ e^{-\beta \Uaug(\bL,\bs) - \beta P V(\bL) + N \log(V(\bL))},\\
        & \hspace{-0.6cm} = N! \int \,d \bL\ \int_{\bs^\dagger}\,d\bs\ e^{-\beta \Uaug(\bL,\bs) - \beta P V(\bL) + N \log(V(\bL))},\\
        & \hspace{-0.6cm} = N!\ \mathbb{E}_{(\bL,\bs) \sim q_X}\left[ \frac{e^{-\beta \Uaug(\bL,\bs) - \beta P V(\bL) + N \log(V(\bL))}}{q_X(\bL,\bs)}\right],
\label{eq:Greweigh2}
\end{aligned}
\end{equation}
where $\bs^\dagger$ denotes one permutation of $\bs$.

Plugging Eq. \ref{eq:Greweigh2} into Eq. \ref{eq:Greweigh1}, we find that the Gibbs free energy of the augmented system can be computed from samples from $q_X$
\begin{equation}
\begin{aligned}
    \beta \Gaug 
    =&  -\log \left(\mathbb{E}_{(\bL,\bs) \sim q_X}\left[e^{-\beta \UGaug(\bL,\bs)} \dfrac{1}{q_X(\bL,\bs)} \right] \right) \\
    & -\log(N!).
\end{aligned}
\end{equation}

The previous equation allows us to compute $\Gaug$ in terms of samples from $q_X$. However, we are interested in the Gibbs free energy of the non-augmented system $G_X$. Therefore, we write the free-energy difference between the augmented and the non-augmented system as
\begin{equation}
    \begin{split}
        \beta \Delta G
        &= \beta G - \beta \Gaug
        = \log \left( \dfrac{\Deltaaug}{\Delta_X} \right)\\
        &\hspace{-1cm}= \log \dfrac{\int \,d \bL\ \int \,d\bs\ e^{-\beta \Uaug(\bL,\bs) - \beta PV(\bL) + N\log V(\bL)}}{\int \,d \bL\ \int \,d\bs\ e^{-\beta U(\bL,\bs) -\beta PV(\bL) +N \log V(\bL)}}  .
        \label{eq:deltaG}
    \end{split}
\end{equation}
Because $U$ is invariant under global translations of the system, it can be rewritten as
\begin{equation}
    U(\bx_1,...,\bx_N) 
    = U(\mathbf{0},\bx'_2,...,\bx'_N) 
    = U(\bx'_2,...,\bx'_N) 
\label{eq:npt_U_invariant}
\end{equation}
where $\bx'_i=\bx_i-\bx_1$ for all $i$ and we remove the first coordinate for the sake of simplicity. Let us now consider a change of variables from $(\bL,\bs_1,\bs_2,...,\bs_N)$ to $(\bL,\scom,\bs_2',...,\bs_N')$, where 
\begin{equation}
    \begin{split}
        \scom 
        &= \dfrac{1}{N} \sum_{i=1}^N \bs_i\\
        \bs_i'
        &= \bs_i - \bs_1.
    \end{split}
\end{equation}

Then, we can substitute $\Uaug(\bL,\bs)$ by $ U(\bL,\bs_2'...,\bs_N') + \Ucom(\scom)$, and $U(\bL,\bs)$ by $U(\bL,\bs_2'...,\bs_N')$ in Eq. \ref{eq:deltaG}, and integrate with respect to $\,d\scom \ d\bs_2' ...\bs_N'$ instead of $d\bs$. We thus simplify Eq. \ref{eq:deltaG} as
\begin{equation}
    \begin{split}
        \beta \Delta G
        &= \log \dfrac{ \int \,d\scom  \exp\left(-\beta \Ucom(\scom)\right)}{\int \,d\scom \ },
    \end{split}
\label{eq:nptlfep_com}
\end{equation}
The two integrals over $\scom$ in Eq. \ref{eq:nptlfep_com} are easy to compute, since one integrand is $1$ and the other integrand is $\exp(-\beta \Ucom(\scom))= \exp(-\beta \kcom \scom^2/2 )$, which is just a Gaussian. However, the exact evaluation of these integrals is beyond the scope of this work.\\

We now have an estimate of the Gibbs free energy of the augmented system via a learned free-energy perturbation (LFEP) approach and an analytic expression for the Gibbs free-energy contribution of the CoM constraint. Therefore, the Gibbs free energy of the non-augmented system can be computed as
\begin{equation}
    \begin{split}
    \beta G_X 
    =& \beta \Gaug + \beta \Delta G_{\text{CoM}}\\
    =&  -\log \left(\mathbb{E}_{(\bL,\bs) \sim q_X}\left[e^{-\beta (\UGaug(\bL,\bs)} \dfrac{1}{q_X(\bL,\bs)} \right] \right) \\
    &\hspace{-1cm}- \log(N!) + \log \dfrac{ \int \,d\scom  \exp\left(-\beta \Ucom(\scom)\right)}{\int \,d\scom \ 1}  .
    \end{split}
\label{eq:npt_lfep}
\end{equation}

Most importantly, when calculating Gibbs free-energy differences, the last two terms in Eq. \ref{eq:npt_lfep}  cancel when the system's number of particles, reference lattice and center-of-mass restraint are kept unchanged.

\subsection{Observables \label{sec:observables}}
In this section, we discuss  observables that can be computed from configurations sampled either by a BG or MD. More precisely, we discuss the radial distribution function and the instantaneous pressure.  These observables will be computed for two case studies in Sections \ref{sec:LJnpt} and Section \ref{sec:hslnpt}. 

\subsubsection{Radial distribution function} \label{sec:rdf}
The radial distribution function (RDF) is a measure of the two-body correlations in liquids and crystals. The RDF $g(r)$  is defined as the ratio between the number density of particles at a certain distance $r$ from a fixed particle and the expected number density at distance $r$ for an ideal gas with the same density. \cite{frenkel_understanding_2002} Numerically, we can compute $g(r)$ for one particle position $\br_i$ in configuration $\bx=(\br_1,...,\br_N)$. To get more statistics, we average over all particles ($i=1,...,N$) and over multiple configurations ($\{\bx_1,...,\bx_B\}$). In the $NPT$ ensemble, configurations are given by $(\bL,\bs)$, so the box size and volume vary with each configuration. Therefore, we compute the RDF as follows
\begin{equation}
    g(r) 
    = \dfrac{1}{B N} \sum_{j=1}^B \sum_{i=1}^N \dfrac{\text{count}(r,i,(\bL,\bs)_j)}{\rho'((\bL,\bs)_j) V_{\text{shell}(r)}},
\label{eq:rdf_npt}
\end{equation}
where $\text{count}(r,i,(\bL,\bs)_j)$ is the number of particles at distance  $[r,r+dr)$ from particle $i$ in configuration $(\bL,\bs)_j$, and $\rho'((\bL,\bs)_j) = (N-1)/V(\bL)$, where $V(\bL)=L_x^2 \lambda$ is the volume of the box in configuration $(\bL,\bs)_j$ and  $V_{\text{shell}}(r)= \pi (r+dr)^2 - \pi r^2 $ is the volume of a circular shell around $\br_i$.

\subsubsection{Pressure}
\label{sec:pressure}
Another quantity we can compute is the instantaneous pressure as defined by the virial equation. For the $NPT$ ensemble, the macroscopic pressure is fixed, because the system is coupled to a (fictitious) barostat at  pressure $P$. The instantaneous pressure $P_{\text{instant}}(\bL,\bs)$ fluctuates around $P$ and, in MD simulations, the volume fluctuates based on the pressure difference $P-P_{\text{instant}}(\bL,\bs)$ between the external and internal system. The instantaneous pressure can be computed via
\begin{equation}
    P_{\text{instant}}(\bL,\bs)
    = \dfrac{N k_B T}{V(\bL)} + \dfrac{\sum_{i=1}^N \mathbf{F}_i \cdot \bx_i}{D V(\bL)},
\end{equation}
where $V(\bL)=L_x^2\lambda$ is the volume of the system, $D$ is the dimensionality of the system and $\mathbf{F}_i=\nabla_{\bx_i}U(\bx_1,...,\bx_N)$ is the force on particle $i$. 

\section{RESULTS}

\subsection{Lennard-Jones system in the $NPT$ ensemble}
\label{sec:LJnpt}

\begin{figure}[h]
    \centering
    \includegraphics[width=0.9\linewidth]{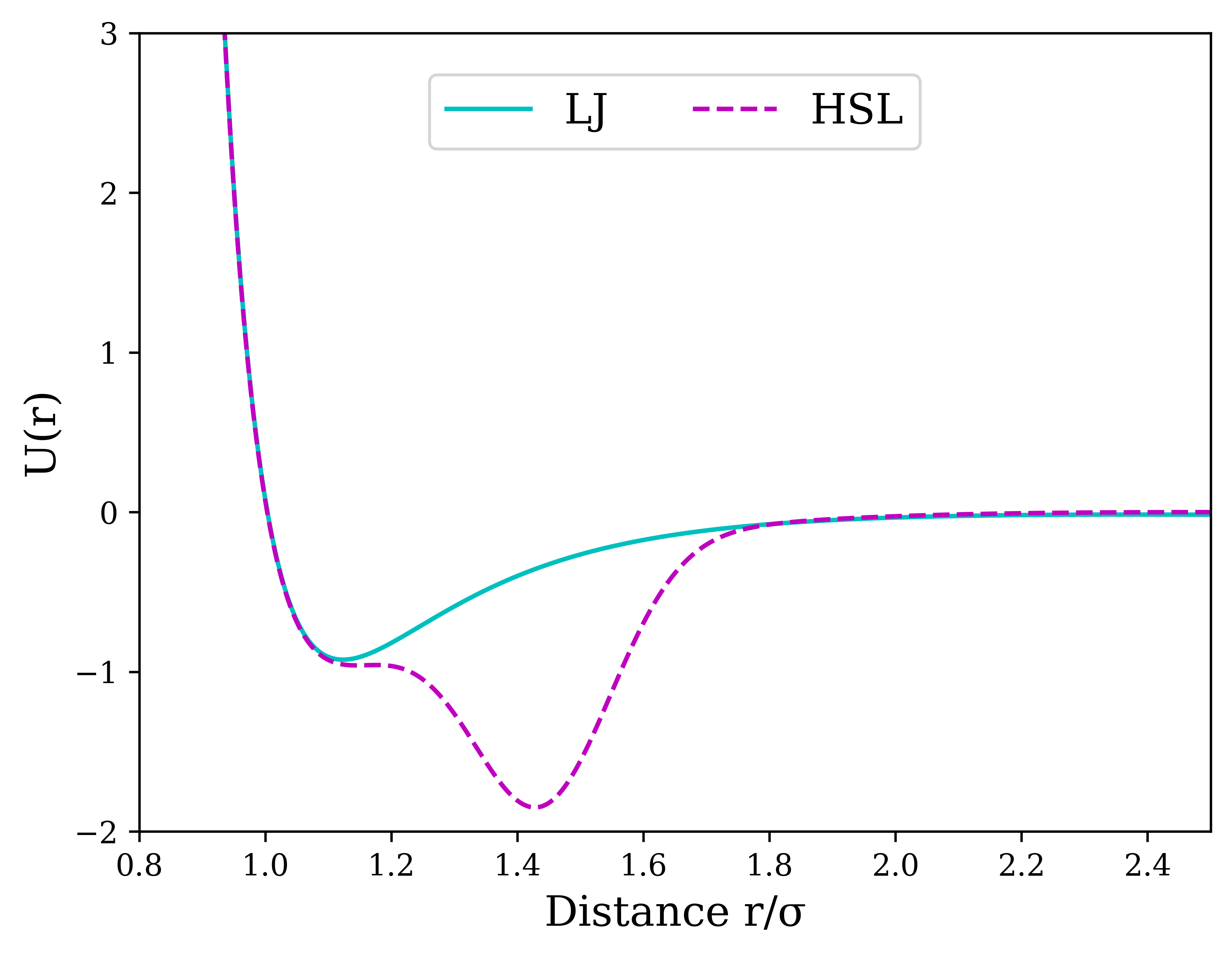}
    \caption{The Lennard-Jones (LJ) and Hemmer-Stell-like (HSL) pair potentials.}
    \label{fig:LJHSL_pot}
\end{figure}

In our first case study, we consider a two-dimensional system with periodic boundary conditions in the $NPT$ ensemble. The system consists of $N=36$ Lennard-Jones (LJ)  particles at pressure $P\sigma^2/\epsilon=5.0$ and temperature $k_B T/\epsilon = 0.5$, where the system has been shown to exhibit a hexagonal crystal phase. \cite{barker_phase_1981} 
The LJ potential is given by
\begin{equation}
    \ULJ(r_{ij})
    = \sum^{N}_{j>i} 4 \epsilon \left[ \left(\frac{\sigma}{r_{ij}}\right)^{12} - \left(\frac{\sigma}{r_{ij}}\right)^{6} \right],\\
\end{equation}
where $r_{ij}$ is the inter-particle distance between particle $i$ and $j$ computed using the nearest image convention. Furthermore, $\sigma$ is the particle diameter and $\epsilon$ is the interaction strength (see Fig. \ref{fig:LJHSL_pot}). We use $\sigma$ and $\epsilon$ as our units of length  and energy, respectively. The potential is truncated and shifted to zero at a chosen cut-off distance $\rcut$. The settings for all these parameters are listed in Table \ref{tab:LJnpt_sys}.

\subsubsection{Boltzmann generator protocol}
We add a harmonic center-of-mass restraint in terms of the deviations $\bs$ to remove the translation invariance. This potential is given by
\begin{equation}
    \Ucom(\bs) 
    = \dfrac{1}{2} \kcom \scom^2
    =\dfrac{1}{2} \kcom \left(\dfrac{1}{N}\sum_{i=1}^N \bs_i \right)^2,
    \label{eq:LJcomres}
\end{equation}
where $\kcom \gg 8 N^2 k_B T$ in order to only sample one permutation (see Eq.  \ref{eq:kcom_min} for a derivation). In this case we have $ N^2 k_B T=8\cdot 36^2 \cdot 0.5 \epsilon=5184\epsilon$. Hence, we set $\kcom=3*10^5 \epsilon$.

Once the translation and permutation are handled, we must also handle numerical instabilities during training. The divergence in $\nabla_r \ULJ(r)$ for $r \downarrow 0$ can lead to instabilities in training the BG. To overcome this problem, we regularize the potential.\cite{wirnsberger_normalizing_2021} We linearize the LJ pair interaction below a distance $\rlin$, such that
\begin{eqnarray}
    \ULJ^\text{reg}(r) && \nonumber \\ 
&&\hspace{-1.5cm}  =  \begin{cases}
        \ULJ(\rlin) + \left(\dfrac{\partial \ULJ}{\partial r}\right)\Bigr|_{r=\rlin} (r - \rlin) & r < \rlin \\
        \ULJ(r) & r \geq \rlin
    \end{cases} .
    \label{eq:LJreg}
\end{eqnarray}

On the one hand $\rlin$ should be chosen far enough away from $r=0$ in order to ensure that $\partial_r \ULJ^\text{reg}(r)$ is small enough. On the other hand, $\rlin$ should be chosen small enough to avoid influencing the generated distribution. More precisely, $\rlin$ should be chosen so that $\ULJ(\rlin) \gg k_B T$, because deviations of the order of $k_B T$ are exponentially unlikely. In practice, we set $\rlin = 0.8 \sigma$ so that
\begin{equation}
    \begin{split}
        \ULJ(r) \Bigr|_{r=\rlin=0.8 \sigma}
        & \approx 49 \epsilon; \\
        \left(\dfrac{\partial \ULJ}{\partial r}\right)\Bigr|_{r=\rlin = 0.8 \sigma}
        & \approx -759 \dfrac{\epsilon}{\sigma}.
    \end{split}
\end{equation}

We train a $NPT$ BG on KL-loss only  to use the BG in an unsupervised
way. We use the RealNVP architecture described in Section \ref{sec:bgs}. We train over several epochs, where each epoch can have  different hyperparameters (e.g. batch size or learning rate). For the specifics about the architecture and hyperparameters used in each epoch, as well as the regularization of the LJ interaction potential, see Section \ref{sec:applj}.

\color{black}
\subsubsection{Molecular dynamics protocol}
We obtain data from MD simulations using LAMMPS.\cite{thompson_lammps_2022} Since the BG is trained on KL-loss only, the MD data is not used as training data, but as the ground truth to compare the BG-generated configurations. The MD data was obtained in the following way. We initialise the particles on a hexagonal lattice. We then simulate for $1 \times 10^6$ MD steps with a time step $\Delta t = 0.001$, saving a configuration every $100$ steps. Furthermore, we remove the first $100$ configurations. Therefore, the MD data consists of $9900$ samples. For more details on the MD simulations, we refer the reader to Section \ref{sec:applj}. 

\FloatBarrier
\subsubsection{Potential energy distribution} \label{sec:LJnpt_hist}
\begin{figure}[h]
    \centering
        \raggedright \hspace{0.1cm} \large (a)\\
        \centering \includegraphics[width=0.8\linewidth]{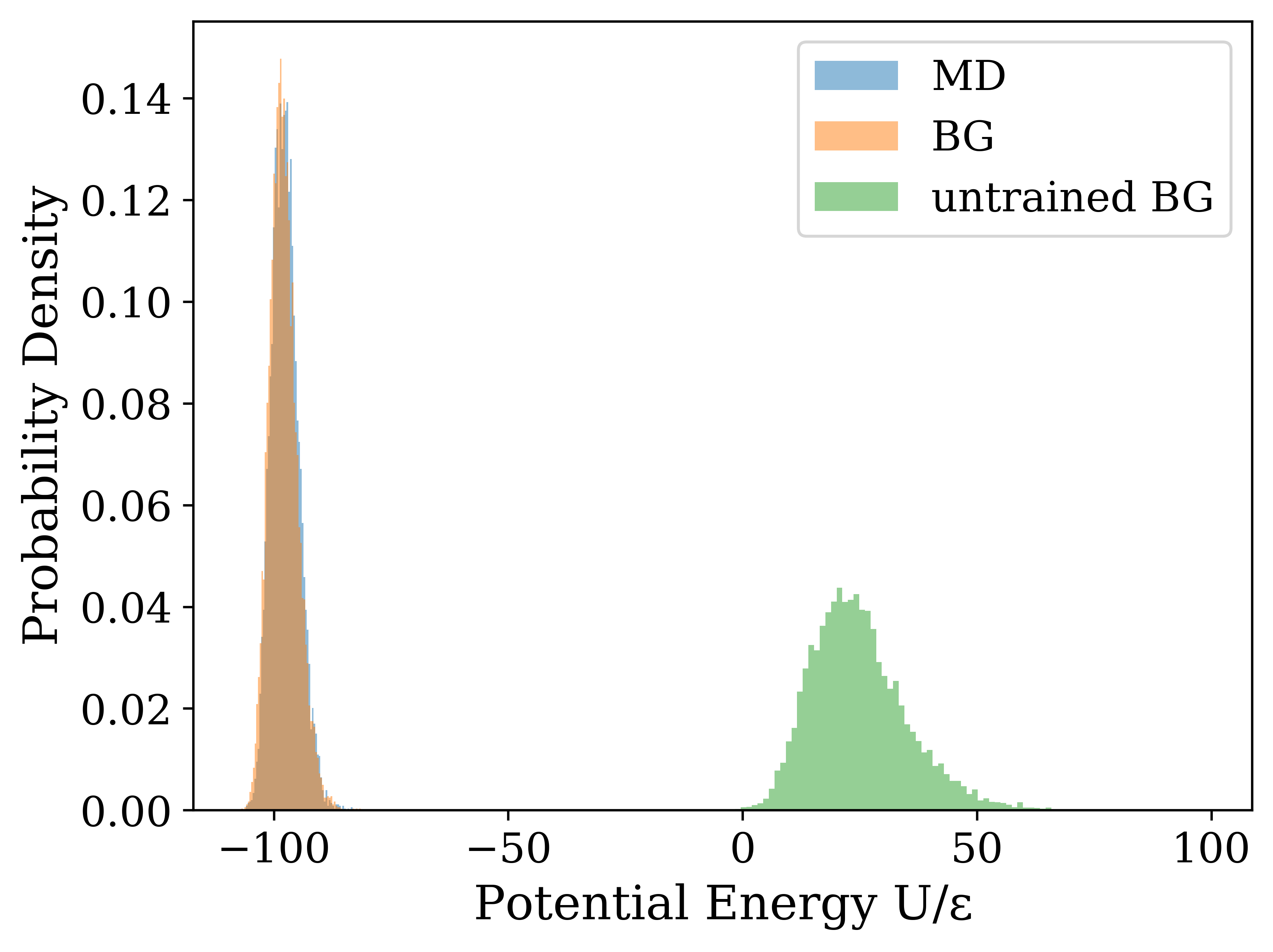}\\
        \raggedright \hspace{0.1cm}(b)\\
        \centering \includegraphics[width=0.8\linewidth]{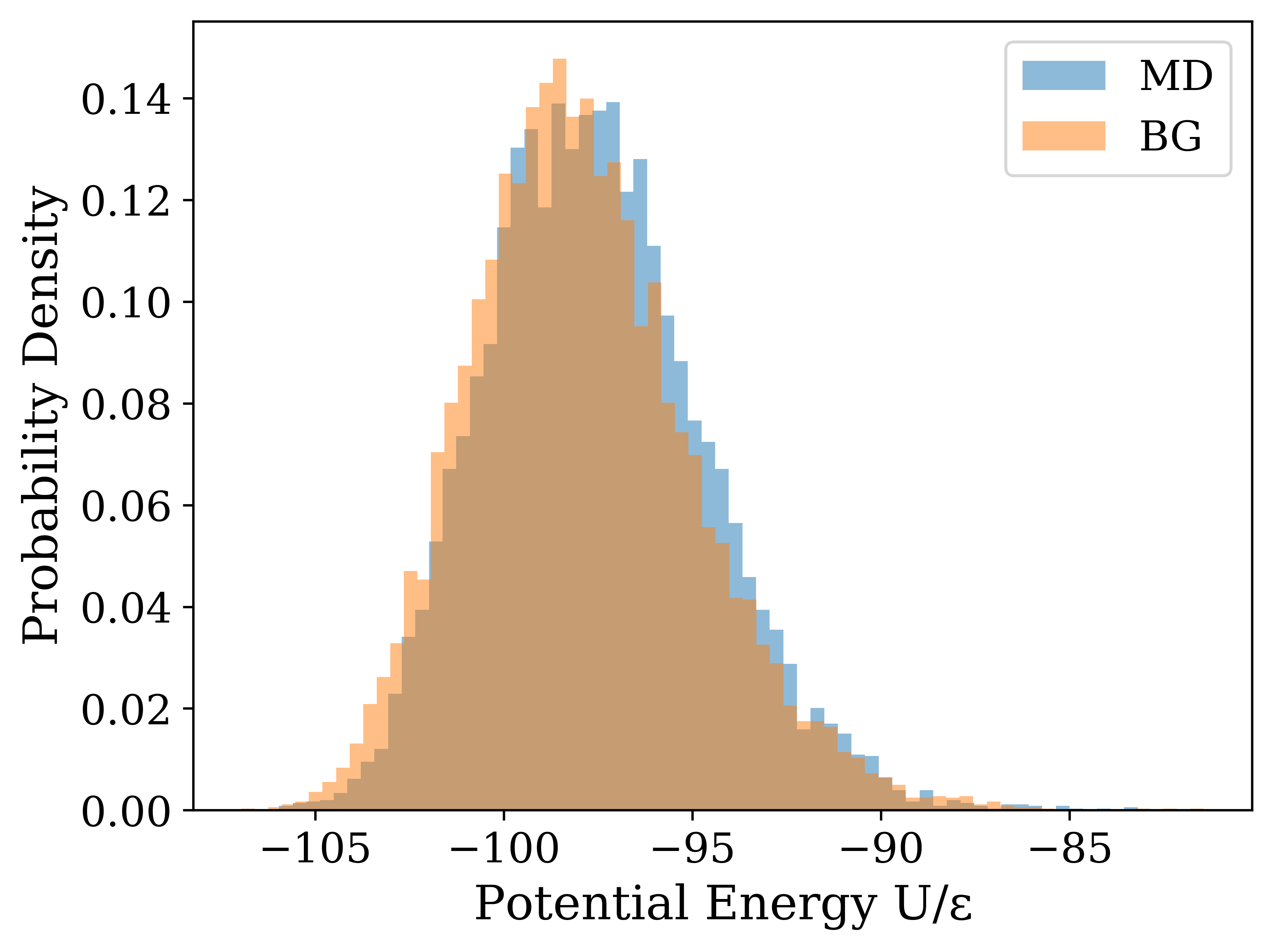}
    \caption{(a) The potential energy distributions corresponding to samples from  MD simulations, an untrained $NPT$ BG, and a trained $NPT$ BG. (b) The  potential energy distributions corresponding to samples from the MD simulations and the trained $NPT$ BG show good agreement. The distributions  are normalized histograms of 30000 $NPT$ BG and 9900 MD samples.}
\label{fig:LJnpt_energyhistogram}
\end{figure}

In this section, we assess the quality of the $NPT$ BG samples by comparing their potential energy distribution to that of the MD samples, which we take to be the ground truth. In Fig. \ref{fig:LJnpt_energyhistogram}, we plot the potential energy distributions obtained from (a) an untrained and (b) a trained $NPT$ BG along with the potential energy distribution from MD simulations.  Fig. \ref{fig:LJnpt_energyhistogram} shows that the trained BG significantly improves upon the untrained BG. Furthermore, Fig. \ref{fig:LJnpt_energyhistogram}(b) reveals that the BG and MD potential energy distributions show good agreement.

\FloatBarrier
\subsubsection{Volume distribution}
\begin{figure}[h]
    \centering
        \raggedright \hspace{0.1cm} \large (a)\\
        \centering\includegraphics[width=0.8\linewidth]{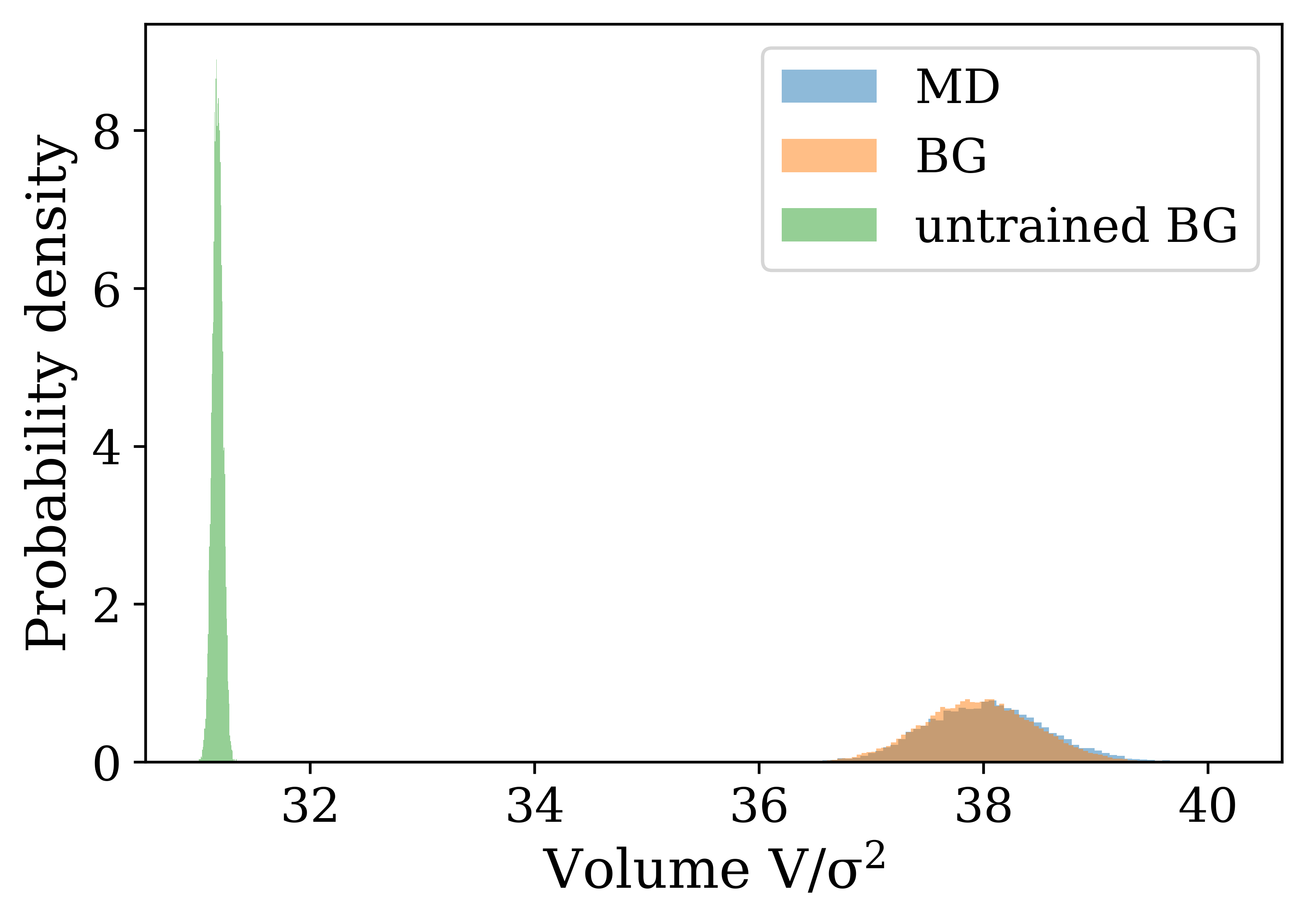}\\
        \raggedright \hspace{0.1cm} \large (b)\\
        \centering\includegraphics[width=0.8\linewidth]{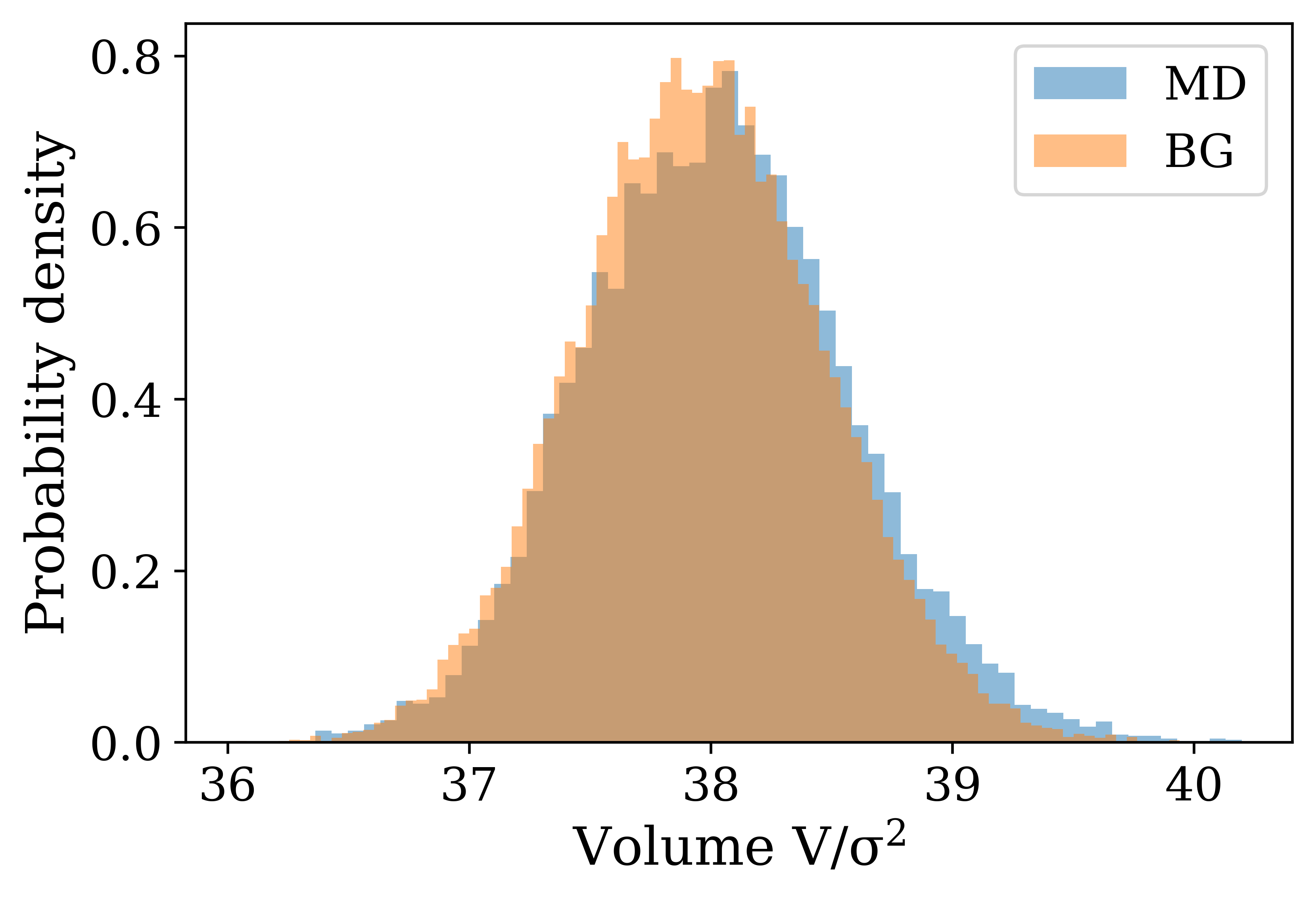}
    \caption{(a) The volume distributions corresponding to samples from MD simulations, an untrained $NPT$ BG, and a trained $NPT$ BG. (b) The potential energy distributions corresponding to samples from MD simulations and the trained $NPT$ BG. The distributions  are normalized histograms of 30000 $NPT$ BG or 9900 MD samples.}
    \label{fig:LJnpt_volume}
\end{figure}

Since the pressure is fixed in the isobaric-isothermal ensemble, the volume is allowed to fluctuate. Fig. \ref{fig:LJnpt_volume}(a) shows the volume distribution of the MD samples versus an untrained and a trained $NPT$ BG. The trained $NPT$ BG clearly improves upon the untrained $NPT$ BG. Furthermore, Fig. \ref{fig:LJnpt_volume}(b) shows just the samples from the trained $NPT$ BG and from MD simulations. The two distributions show good agreement. The $NPT$ BG slightly undersamples the larger volumes.

\FloatBarrier
\subsubsection{Pressure distribution}
\begin{figure}[h]
    \centering
    \includegraphics[width=0.8\linewidth]{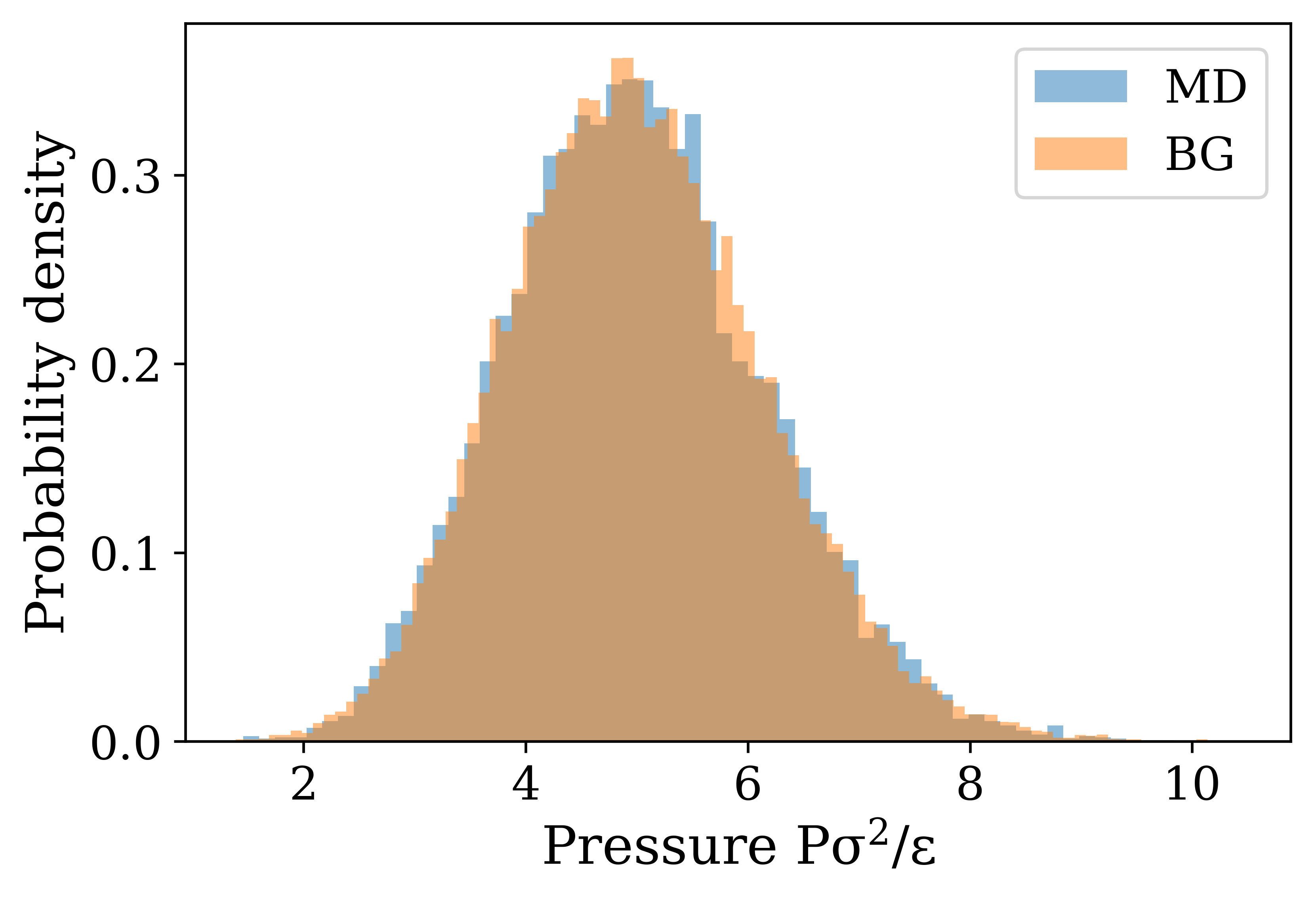}
    \caption{The instantaneous pressure distributions corresponding to samples from MD simulations and the trained $NPT$ BG. The distributions are normalized histograms of 30000 $NPT$ BG or 9900 MD samples.}
    \label{fig:LJnpt_pressure}
\end{figure}

Subsequently, we compare the instantaneous pressure of MD and $NPT$ BG generated configurations (see Section \ref{sec:pressure} for a definition of the instantaneous pressure). Fig.  \ref{fig:LJnpt_pressure} shows that the distribution of the pressure of the $NPT$ BG generated configurations has good overlap with the pressure distribution of configurations from MD simulations. Furthermore, both distributions are centred around the macroscopic pressure $P\sigma^2/\epsilon=5.0$, as expected.

\FloatBarrier
\subsubsection{Radial distribution function} \label{sec:LJnpt_rdf}
\begin{figure}[h]
    \centering
    \includegraphics[width=0.9\linewidth]{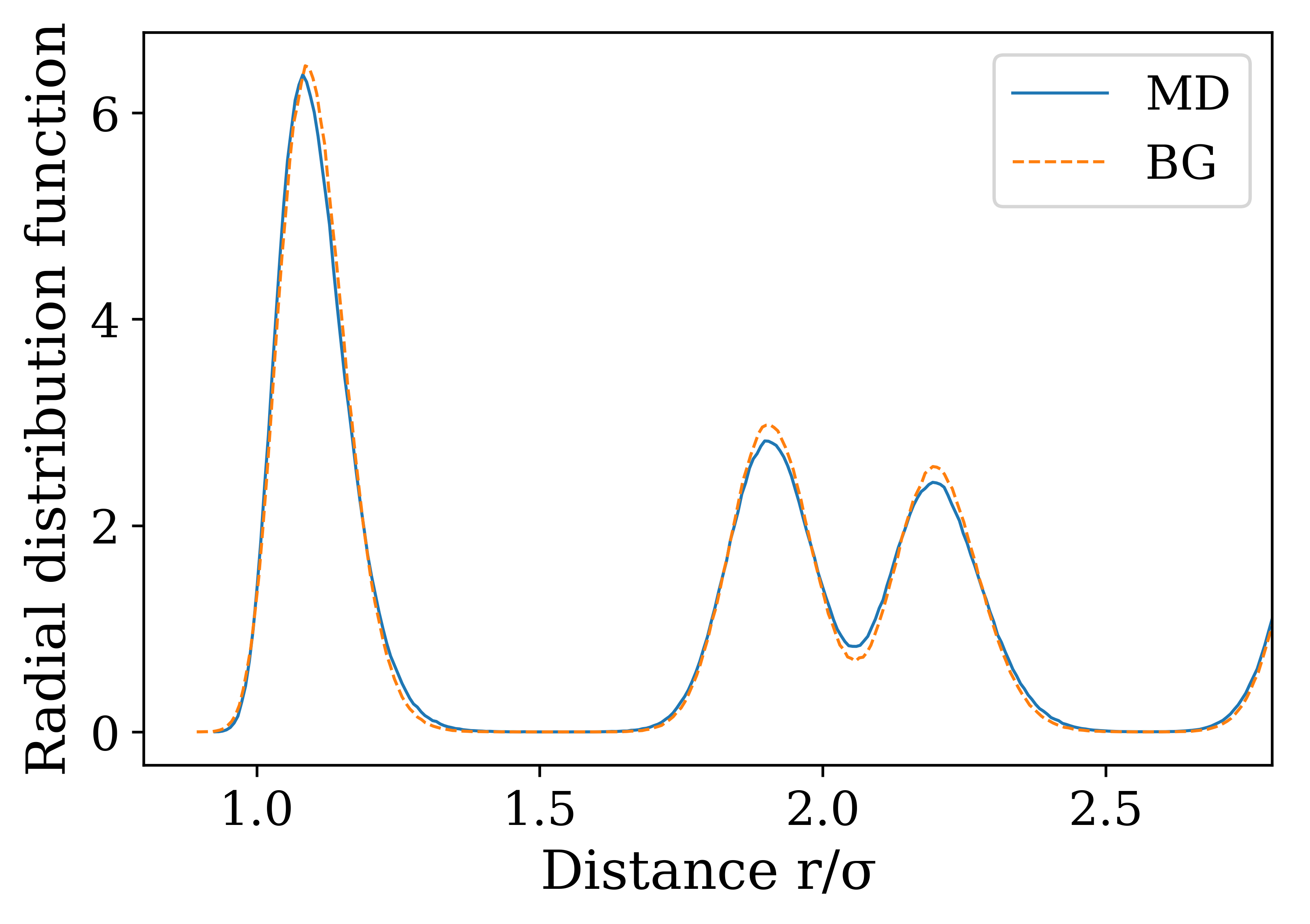}
    \caption{The radial distribution function corresponding to samples from MD simulations and to samples generated by the trained BG. The distributions are generated using 30000 BG or 9900 MD samples.}
    \label{fig:LJnpt_rdf}
\end{figure}

Finally, we examine the radial distribution functions as obtained by the $NPT$ BG. Figure \ref{fig:LJnpt_rdf} shows that the BG radial distribution function (RDF) has peaks at the same distances as the MD one. This indicates that the BG samples the hexagonal crystal structure properly. However, the peaks of the BG RDF are slightly narrower and higher. This could indicate that the BG generates deviations that are closer to the lattice sites. Reweighting the BG generated samples could remove this discrepancy\cite{noe_boltzmann_2019}.

\subsection{Hemmer-Stell-like system in the $NPT$ ensemble}
\label{sec:hslnpt}
We now turn our attention to the second case study, which is a Hemmer-Stell-like (HSL) system as discussed in Section \ref{sec:apphsl}. 
We  consider  a 2D system with periodic boundary conditions in the $NPT$ ensemble. The system consists of $N=36$ particles interacting with an HSL potential\cite{quigley_progression_2005} at a temperature $k_B T/\epsilon = 0.1$. This system has  shown to exhibit a metastable isostructural phase transition.\cite{quigley_progression_2005} We train two $NPT$ BGs, one at pressure $P\sigma^2/\epsilon=5.0$ and the other at pressure $P\sigma^2/\epsilon=2.0$.  The HSL potential is given by
\begin{eqnarray}
    U_\text{HSL}(r)
    &=& 4\epsilon \left[ \left(\frac{\sigma}{r}\right)^{12} - \left(\frac{\sigma}{r}\right)^{6} \right]\nonumber \\
    &-&   A \exp[-w (r_{ij}- r_0)^2], 
\end{eqnarray}
where the first term is the same as in the LJ system, and the second term corresponds to an exponential well with parameters $A$ and $w$, centered at $r_0$\cite{quigley_progression_2005} (see Fig. \ref{fig:LJHSL_pot}). The settings for these parameters are listed in Table \ref{tab:HSLnpt_sys}.

\subsubsection{Boltzmann generator protocol}
To sample only one permutation and remove translation invariance, we add a harmonic center-of-mass restraint in terms of the deviations $\bs$ with the same parameters as for the LJ system (see Eq \ref{eq:LJcomres}). Similarly, to avoid numerical instabilities during training, we apply the same regularization strategy as we used for the LJ system (see Eq. \ref{eq:LJreg}). 

\color{black}
\subsubsection{Molecular dynamics protocol}
We again collect data from MD simulations using LAMMPS.\cite{thompson_lammps_2022}  The lattice initialization and the number of MD samples is the same as in the LJ case. For additional details on the MD simulations, we refer to Section \ref{sec:apphsl}. 

 The BG is again trained on KL-loss only, which means that the MD data is solely used for comparison. We also use the same RealNVP architecture as for the LJ system, described in Section \ref{sec:bgs}. Specifics about the architecture and training hyperparameters can be found in Section \ref{sec:apphsl}.

\subsubsection{Potential energy distribution} \label{sec:HSLnpt_hist}

\begin{figure}[h]
    \centering
\includegraphics[width=0.9\linewidth]{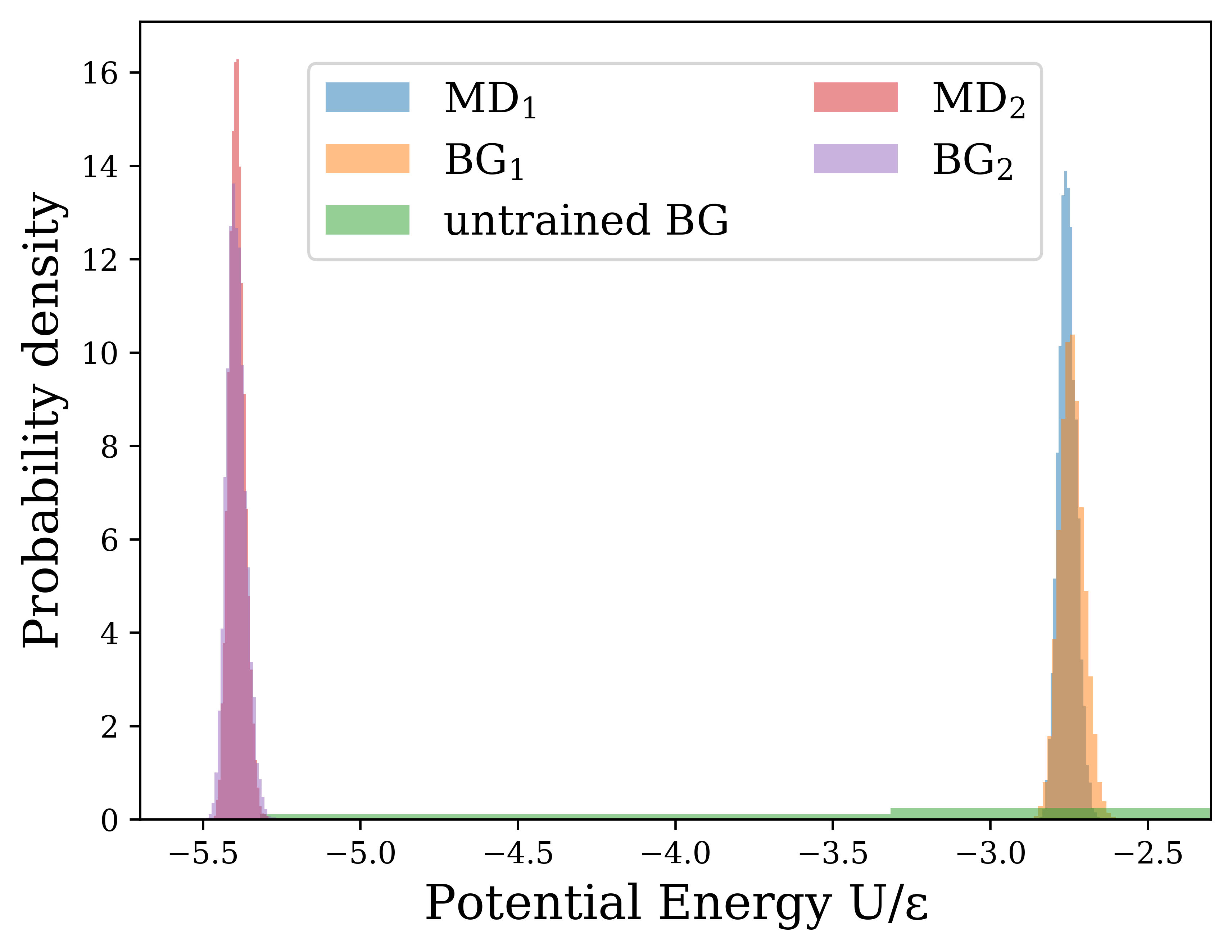}
    \caption{The potential energy distributions corresponding to samples from MD simulations, an untrained $NPT$ BG, and a trained $NPT$ BG at $P\sigma^2/\epsilon=5.0$ (MD$_1$ and BG$_1$) and $P\sigma^2/\epsilon=2.0$ (MD$_2$ and BG$_2$). The distributions are generated using 10000 BG or 9900 MD samples.}
    \label{fig:HSLnpt_energy}
\end{figure}

In Fig. \ref{fig:HSLnpt_energy} we present the potential energy distributions obtained from  MD samples versus those of an untrained $NPT$  BG (untrained BG) and a trained BG for both the high-pressure (MD$_1$ and BG$_1$) and the low-pressure  (MD$_2$ and BG$_2$) HSL systems. We clearly observe that the potential energy distributions obtained from the trained BGs improve considerably with respect to the one from the untrained BG. For both pressures,  the potential energy distributions from the trained BGs are centered at approximately the same value as their MD counterparts. Nonetheless, the high-pressure BG shows slightly higher potential energies than the corresponding MD sampling. This is likely because, by being anchored to the reference lattice, the BG sampling fluctuates less into the exponential well of the HSL potential (see Fig. \ref{fig:LJHSL_pot}). This mismatch can be solved by reweighting the BG samples according to their Boltzmann weight.\cite{noe_boltzmann_2019}

\FloatBarrier
\subsubsection{Volume distribution}

\begin{figure}[h]
    \centering
\includegraphics[width=0.9\linewidth]{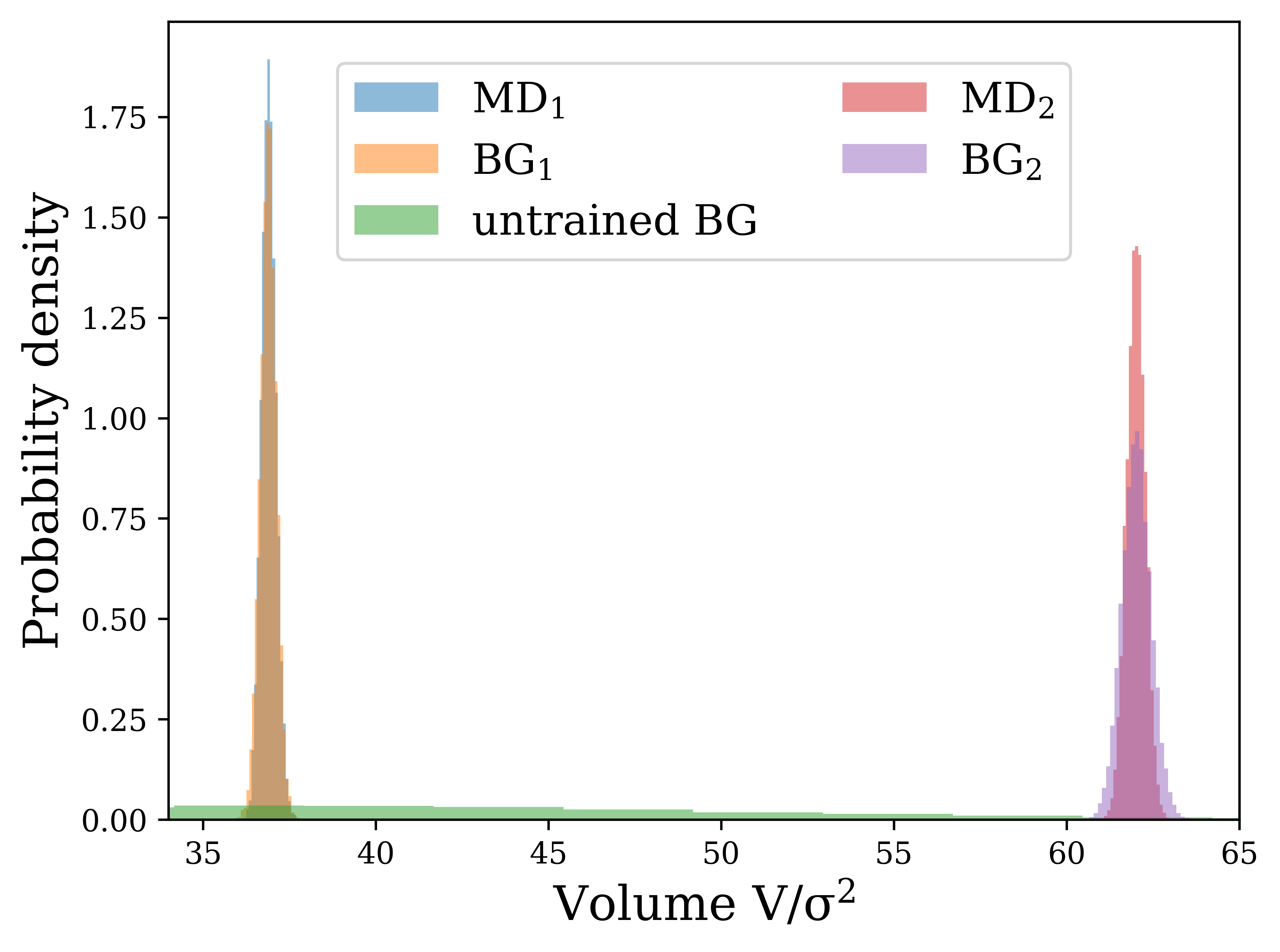}
    \caption{The volume distributions corresponding to samples from MD simulations and the $NPT$ BG at $P\sigma^2/\epsilon=5.0$ (MD$_1$ and BG$_1$) and $P\sigma^2/\epsilon=2.0$ (MD$_2$ and BG$_2$), as well as an untrained BG. The distributions are generated using 10000 BG or 9900 MD samples.}
    \label{fig:HSLnpt_volume}
\end{figure}

Fig. \ref{fig:HSLnpt_volume} presents the volume distribution of MD samples versus those of an untrained $NPT$ BG (untrained BG) and a trained $NPT$ BG  for both the high-pressure (MD$_1$ and BG$_1$) and the low-pressure  (MD$_2$ and BG$_2$) cases. Again the volume distributions improve significantly upon training the BG. The volume distributions from the trained BG are centered at the same value as their corresponding MD distributions for both pressures. However, the high-pressure BG distribution shows better overlap with its MD counterpart compared to the low-pressure one, i.e. the low-pressure BG distribution is not as peaked as its MD equivalent. This is likely due to the larger fluctuations from the reference lattice that occur at lower pressures. After this observation, the number of training epochs of the low-pressure BG was increased (see Section \ref{sec:apphsl}). However, the improvement was marginal. Since the distributions are considerably close, one could still reweight the BG samples to improve the match to MD.

\FloatBarrier
\subsubsection{Pressure distribution}

\begin{figure}[h]
    \centering
\includegraphics[width=0.9\linewidth]{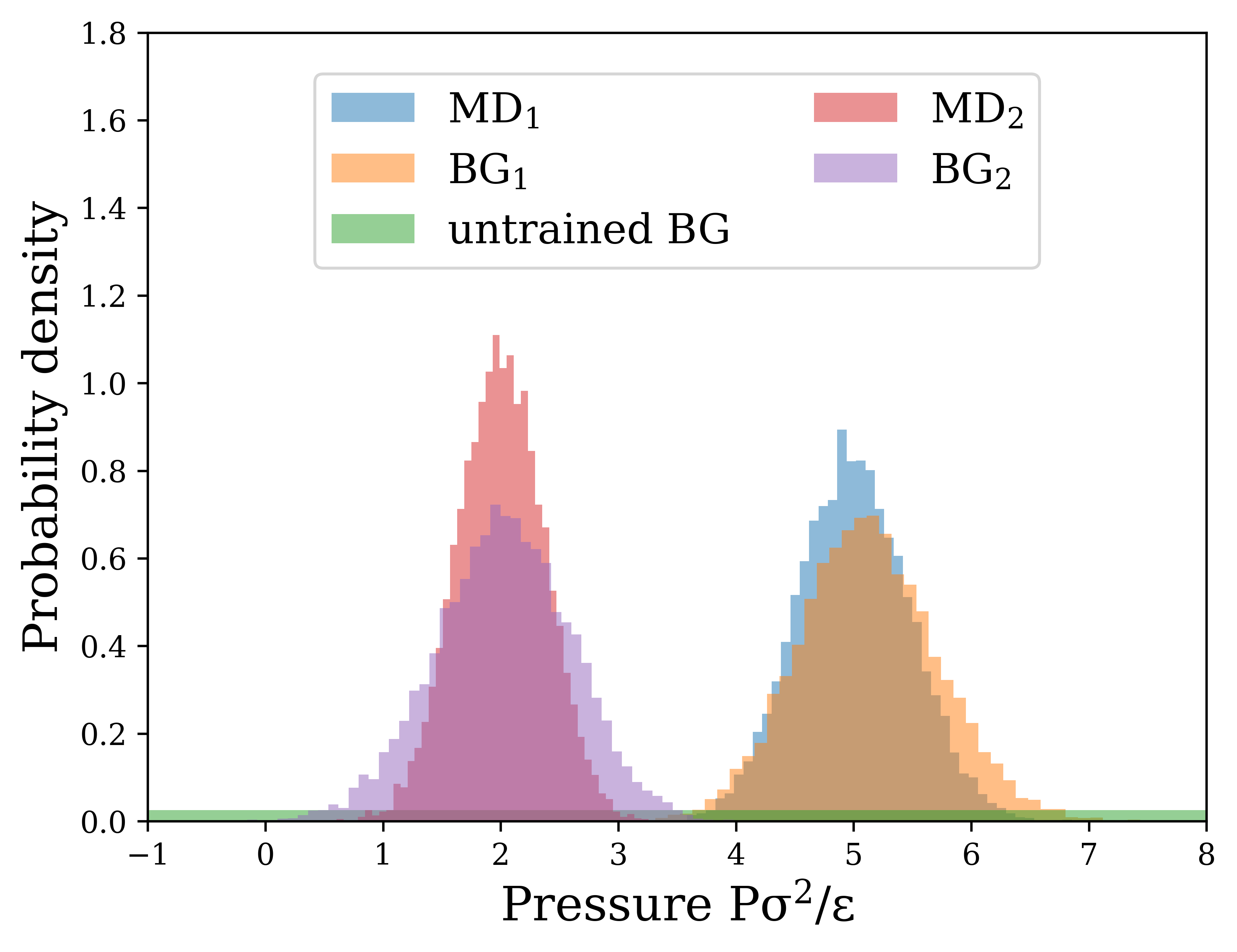}
    \caption{The pressure distributions corresponding to samples from MD simulations and the $NPT$ BG at $P\sigma^2/\epsilon=5.0$ (MD$_1$ and BG$_1$) and $P\sigma^2/\epsilon=2.0$ (MD$_2$ and BG$_2$), as well as an untrained BG. The distributions are generated using 10000 BG or 9900 MD samples.}
    \label{fig:HSLnpt_pressure}
\end{figure}

Fig. \ref{fig:HSLnpt_pressure} shows the instantaneous pressure distribution of MD samples versus BG samples for both the high-pressure (MD$_1$ and BG$_1$) and the low-pressure (MD$_2$ and BG$_2$) systems as well as the distribution of the untrained BG. Even though the instantaneous pressure distributions from the trained BG show reasonable overlap with those obtained from the MD samples, the low-pressure BG distribution is not as peaked as its MD counterpart, similarly to the volume distributions (see Fig. \ref{fig:HSLnpt_volume}). 
On the other hand, the high-pressure BG samples slightly higher pressures than its MD counterpart, which is also reflected in the higher potential energy (see Fig. \ref{fig:HSLnpt_energy}). It is also worth noting that the BG pressure distributions are considerably similar to each other in height and width, with only the mean changing from $P\sigma^2/\epsilon=2.0$ to $5.0$, whereas the MD distributions change significantly. This might indicate that the NF mainly performs a translation of the pressure distribution.

\FloatBarrier
\subsubsection{Radial distribution function}
\begin{figure}
   \centering
        \raggedright \hspace{0.1cm} \large (a)\\
        \centering\includegraphics[width=0.9\linewidth]{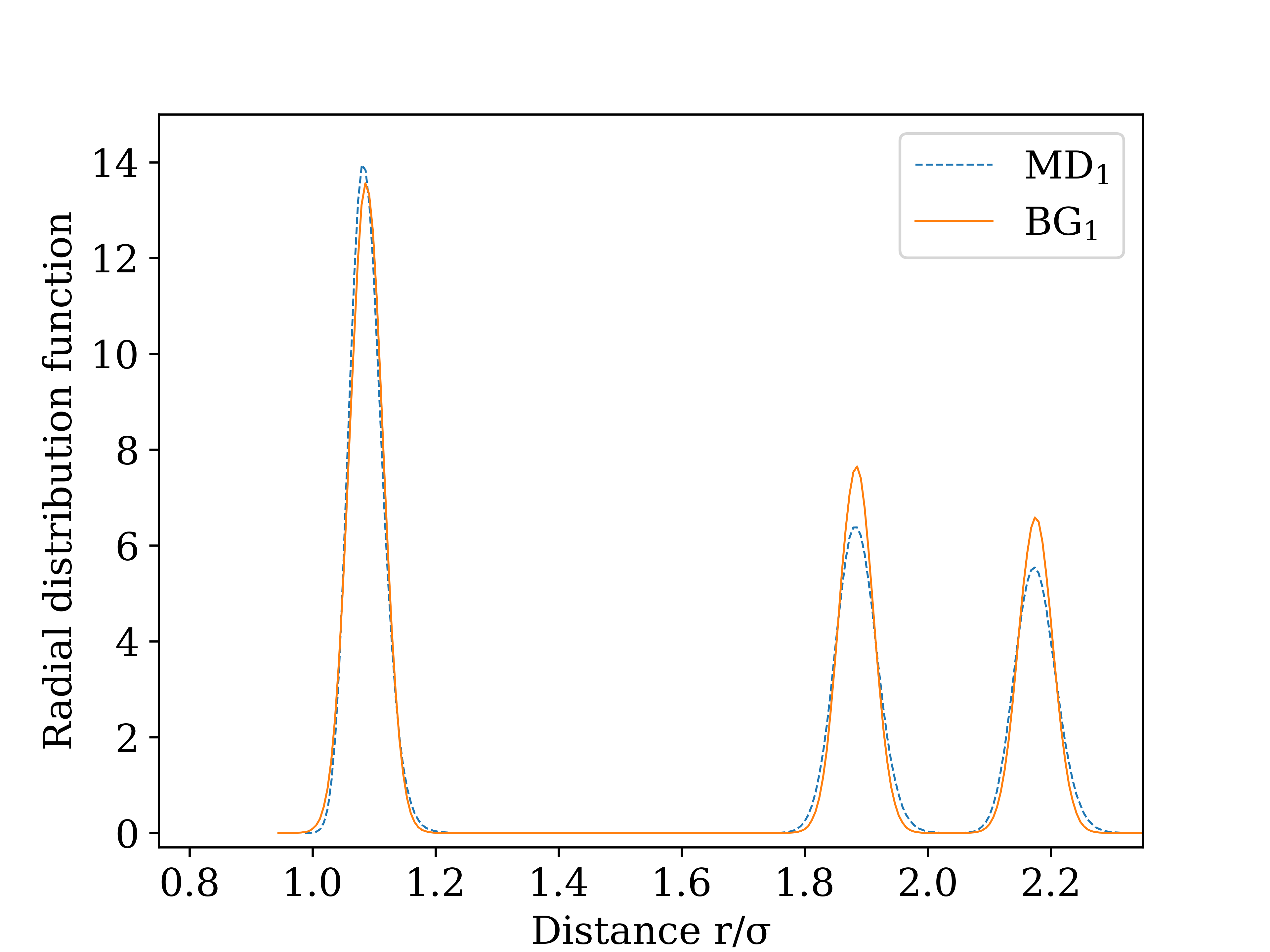}\\
        \raggedright \hspace{0.1cm} \large (b)\\
        \centering\includegraphics[width=0.9\linewidth]{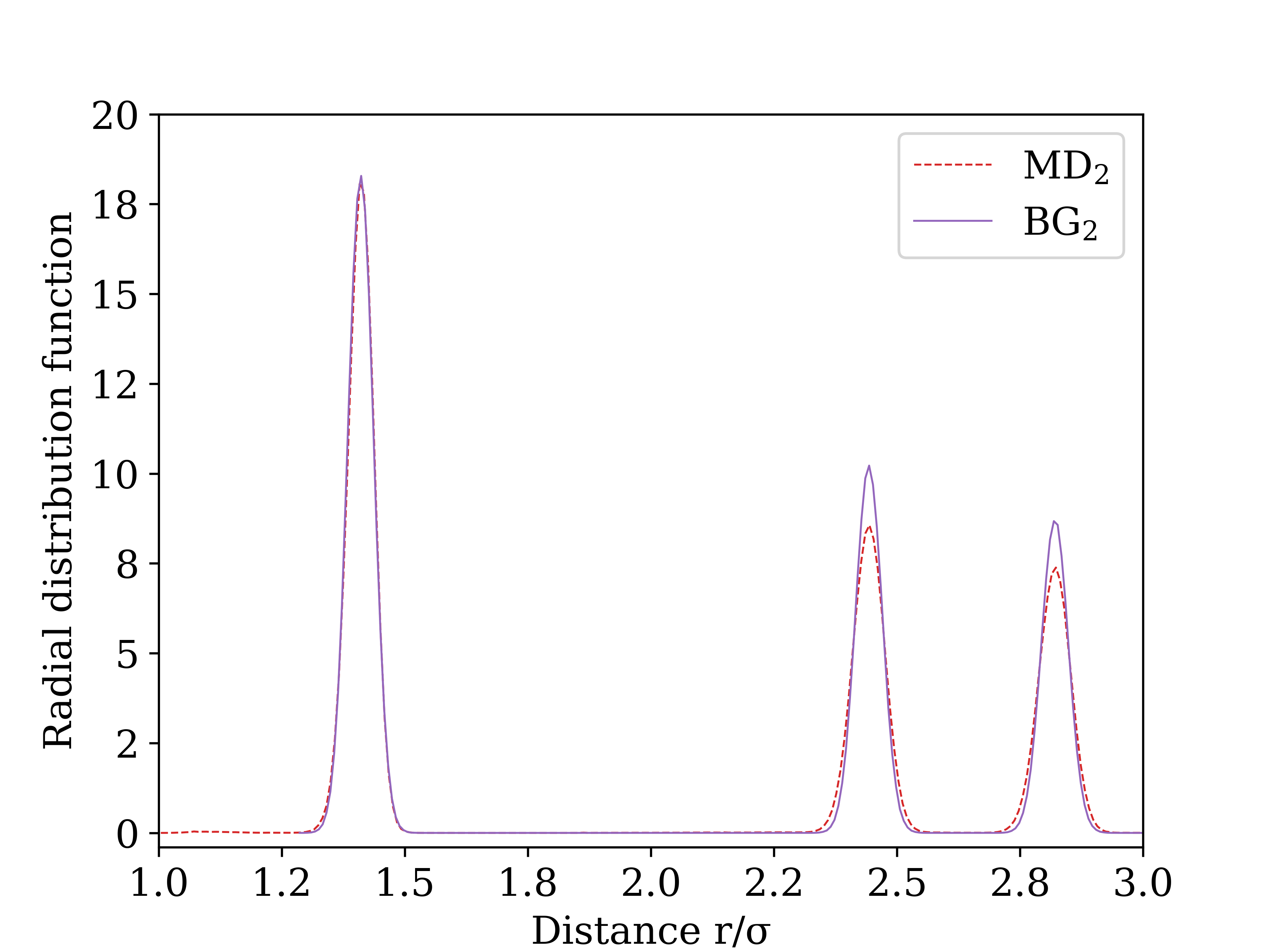}
    \caption{(a) The radial distribution function (RDF) at high pressure, i.e. $P\sigma^2/\epsilon=5.0$, corresponding to samples from MD simulations and to samples generated by the BG. The distributions are generated using 10000 BG or 9900 MD samples. (b) The RDF at low pressure, i.e. $P\sigma^2/\epsilon=2.0$, corresponding to samples from MD simulations and to samples generated by the BG. The distributions are generated using 10000 BG or 9900 MD samples.}
    \label{fig:HSLnpt_rdf}
\end{figure}

In Fig. \ref{fig:HSLnpt_rdf} we show the RDF for (a) the high-pressure BG and (b) the low pressure BG. Both are compared to the corresponding MD RDF. In both cases, the consistent distance between peaks in the BG and the MD RDF shows that the hexagonal phase is sampled properly. Moreover, the similarity between the high-pressure and low-pressure RDFs also confirms that we sample the high- and low-density hexagonal phases. Similar to the LJ system, the BG RDFs are slightly more peaked than their MD counterparts. Again, this is a consequence of the underlying reference lattice, from which the BG generates deviations.

\color{black}
\FloatBarrier
\section{CONCLUSION}

In this work, we presented and tested the first BG for the $NPT$ ensemble, which is based on previous NFs for crystal structures.  \cite{ahmad_free_2022,wirnsberger_normalizing_2021} The key idea behind the $NPT$ BG is to generate not only particle positions, but also fluctuations of the box itself. This adds one other dimension to the configuration space $X$ and latent space $Z$ of the NF, requiring some methodological adjustments as described in Section \ref{sec:npt_bg}. We used the new $NPT$ BG to generate unsupervised one-shot configurations according to the isobaric-isothermal probability distribution. These samples show good agreement with configurations from MD simulations, as demonstrated for the LJ system.  Moreover, we  show that the $NPT$ BG can sample low-density and high-density  isostructural phases, as exemplified for the HSL system. Additionally, we derived an expression of the Gibbs free energy in terms of samples generated by the $NPT$ BG.

While the BG  samples in the $NPT$ ensemble have good agreement---especially regarding mean values---with samples from MD, there are a few differences to keep in mind. The main difference is observed in the distributions of quantities such as volume or instantaneous pressure, or RDFs, which have narrower and higher peaks in BG sampling than in MD. This is due to the underlying reference lattice approach.\cite{ahmad_free_2022} While this approach eliminates the computationally prohibitive cost of sampling multiple permutations, it also anchors the sampling to a fixed reference, hence leading to narrower sampling. Increasing training does not seem to significantly improve the narrow sampling. One possible solution for this is to simply reweight BG samples by their Boltzmann weight. This is feasible because the BG distributions already have good overlap with the MD distributions. Another possible solution is to have a BG with a permutation invariant NF, which is further discussed below. Another difference between BG and MD sampling is observed in the instantaneous pressure distributions, for which the BG shows very similar distributions in height and width for two different pressure values, implying that it learns a translation of the pressure distribution.
\color{black}

 The $NPT$ BG presented in this work opens up exciting possibilities for future research. One possible application is in the screening of pressure-driven transitions or Gibbs free-energy differences in soft materials or atomic systems without relying on computationally expensive MC or MD simulations. Furthermore, the $NPT$ BG  can also be used to investigate various molecular or colloidal processes where pressure plays a critical role. 

Several possible extensions to our proposed method can be considered. An interesting potential improvement is to introduce  anisotropic box fluctuations, where $\lambda$ in $L_y = \lambda L_x$ is not constant. This can be achieved by adding either $\lambda$ or $L_y$  to the generated configuration space $X$. Similarly, this idea can also be extended to generate the angles between the box vectors. Furthermore, the $NPT$ BG can be applied to 3D systems  with similar  flexibility. These extensions will be explored in  future work. 

One limitation of our approach is that it relies on a single reference lattice. Although this allows us to avoid  the combinatorial explosion associated with permutational symmetry, it is also important to consider  transitions between two different lattice types. We discuss two potential solutions to this problem.

One solution to overcome the limitation of our approach based on a single reference lattice is to extend another type of BG using equivariant NFs \cite{wirnsberger_normalizing_2021, wirnsberger_targeted_2020, satorras_en_2021, kohler_equivariant_2020} from the $NVT$ ensemble to the $NPT$ ensemble. These BGs use a prior distribution $\mu_Z$ that is invariant under a group $G$, such as the permutation group, \cite{wirnsberger_targeted_2020, wirnsberger_normalizing_2021}  and an NF that preserves this invariance, such that the generated distribution $q_X$ is  invariant under $G$. This solves the problem between BGs and permutation symmetry. However, a configuration in the $NPT$ ensemble is given by $(\bL, \bs_1,...,\bs_N)$, but the permutation invariance only applies to $(\bs_1,...,\bs_N)$ and not to $\bL$. Hence, the challenge when extending these BGs to the $NPT$ ensemble is to construct an NF that is  equivariant with respect to the last $N*D$ coordinates alone.

The second solution we propose is to use two separate BGs rather than using  one BG to sample both phases simultaneously. A similar method was proposed by No\'{e} {\em et al.}. \cite{noe_boltzmann_2019} Specifically,  two BGs (BG1 and BG2) can be trained with two different reference lattices to sample each of the two crystal phases.  
Using Eq. \ref{eq:npt_lfep}, the absolute Gibb free energy for each phase can then be computed separately.  The Gibbs free-energy difference between the two phases can be obtained by subtracting these two Gibbs free energies. While this method does not provide information about the free-energy barrier, it is  a powerful tool for quickly comparing the Gibbs free energies of various phases. Additionally, this method can be extended  to obtain free-energy differences between arbitrary combinations of interaction potentials and lattices, opening up interesting possibilities for quickly screening free energies of soft materials as the BG only requires modest training. Here we have focused on training by KL-loss to use the BG in an unsupervised manner. However ML training could also be used, even including examples from phases different from the lattice. Pretraining models with ML-loss can help to ensure that BGs show a competitive advantage over MC or MD simulations in terms of the number of energy evaluations.

As a final outlook, the current $NPT$ BG could be integrated with other existing methods for $NVT$ BGs, such as reaction coordinate (RC) loss \cite{noe_boltzmann_2019} and MC moves in latent space. \cite{gabrie_adaptive_2022} The RC loss can bias the BG to produce more samples along a given transition descriptor. This loss could be used for the $NPT$ BG to generate more samples along a phase transition. Furthermore, NFs have been used to generate efficient update moves for MC simulations by performing MC moves in latent space. \cite{gabrie_adaptive_2022} This method could be extended from the $NVT$ ensemble to the $NPT$ ensemble by using an $NPT$ NF. Other advancements in BGs that could be combined with the $NPT$ version include annealed importance sampling, \cite{midgley2022flow} temperature-steerable flows, \cite{dibak2022temperature} conditioning for rare events, \cite{falkner2022conditioning} and diffusion models. \cite{jing2022torsional}

\section*{ACKNOWLEDGEMENTS}
We acknowledge financial support from the European Research Council (ERC Advanced Grant No. ERC-2019-ADV-H2020 884902, SoftML).

\section*{DATA AVAILABILITY}

The code for the $NPT$ BG will be made available in a public GitHub repository at \url{https://github.com/MarjoleinDijkstraGroupUU}. 

\appendix
\section{APPENDIX}

\color{black}
\label{sec:app}

\subsection{Lennard-Jones system in the $NPT$ ensemble}
\label{sec:applj}

In this section, we present details on the generation and training of the LJ system with periodic boundary conditions in the $NPT$ ensemble. More specifically, we provide the parameters for the LJ model and the MD simulations, as well as the architecture and training schedule of the BG.

\subsubsection{Boltzmann generator protocol}

The parameters for the LJ potential, its cut-off distance, regularization and the center-of-mass restraint are specified in Table \ref{tab:LJnpt_sys}.\color{black}

\begin{table}[h!]
    \centering
    \begin{tabular}{c|c|c|c|c|c|c|c}
    $\sigma$ & $\epsilon$ & $N$ & $P$ & $k_B T$ & $\rlin$ & $r_{cut}$ & $\kcom$ \\
    \hline 
    1.0 & 1.0 & 36 & 5.0 & 0.5 & 0.8 & 2.5 & $3*10^5$
    \end{tabular}
    \caption{Model parameters for the $NPT$ LJ system. All values are in reduced units $\sigma$ and $\epsilon$, which are the particle diameter and the interaction strength, respectively.}
    \label{tab:LJnpt_sys}
\end{table}

We train an $NPT$ BG on KL loss only. The architecture of the BG is discussed in Section \ref{sec:bgs} and \ref{sec:npt_bg}. In Table \ref{tab:LJnpt_BG} we specify the hyperparameters used in this architecture. Furthermore, the BG is trained over three epochs, where the batch size is the same in each epoch and the learning rate is gradually decreased over the epochs (see Table \ref{tab:LJnpt_schedule}). \\

\begin{table}[h!]
    \centering
    \begin{tabular}{c|c|c|c|c}
    $\nblocks$ & $\nlayers$ & $\nnodes$ & $C_{\text{lin}}^{\text{init}}$ & $C_{\text{scale}}^\text{init}$ \\
    \hline 
    12 & 3 & 300 & 6.0 & $1/(36*v_{\text{max}})$\\
    \end{tabular}
    \caption{The parameters of the $NPT$ BG, where $\nblocks$ is the number of RealNVP blocks, $\nlayers$ is the number of hidden layers in the neural networks $S$ and $T$ (Section \ref{sec:bgs} explains how $S$ and $T$ are used in the BG), and $\nnodes$ is the number of nodes for each hidden layer in $S$ and $T$. Furthermore, $C_{\text{lin}}^{\text{init}}$ and $C_{\text{scale}}^\text{init}$ are the initial settings of $C_{\text{scale}}$ and $C_{\text{lin}}$ (see Section \ref{sec:npt_bg}).}
    \label{tab:LJnpt_BG}
\end{table}

\begin{table}[h!]
    \centering
    \begin{tabular}{c||c|c|c}
    epoch & 1 & 2 & 3  \\
    \hline 
    \hline iter & 4000 & 4000 & 2000\\
    \hline batch & 256 & 256 & 256  \\
    \hline lr & $10^{-3}$ & $10^{-4}$ & $10^{-5}$\\
    \hline $w_\text{M L}$ & 0 & 0 & 0 \\
    \hline $w_\text{K L}$ & 1 & 1 & 1
    \end{tabular}
    \caption{Training schedule for the $NPT$ BG for the LJ system, where iter is the number of iterations used for training, batch is the batch size, lr is the learning rate used in the Adam optimiser and $w_\text{ML}$ and $w_\text{KL}$ are the weights for the ML and KL loss, respectively. The schedule consists of 3 epochs. Each epoch has different hyperparameters. }
    \label{tab:LJnpt_schedule}
\end{table}

\FloatBarrier
\subsubsection{Molecular dynamics protocol}

We generate MD data by performing simulations using LAMMPS.  \cite{thompson_lammps_2022} Note that this data is not used for training the BG, because we only train on KL loss. It is used only for comparison. We initialise the particles on a hexagonal lattice. This lattice has unit vectors $\ba_1 = (1,0)$ and $\ba_2 = (0,\sqrt{3})$. There are two atoms within this unit cell. We repeat this unit cell 6 times in the $x$-direction and 3 times in the $y$-direction (this gives $6\times3\times2=36$ particles). After initialising the system, we simulate for $10^6$ MD steps with time step $\Delta t = 0.001$, saving a configuration every $100$ steps. Furthermore, we remove the first $100$ configurations. Therefore, the MD data consists of $9900$ samples. We use a Nos\'{e}-Hoover thermostat\cite{martyna_nosehoover_1992} and barostat\cite{parrinello_polymorphic_1981} to ensure constant temperature and pressure, respectively. We use a time constant of $\Delta t_T = 100 \Delta t$ for the thermostat and a time constant of $\Delta t_P = 1000 \Delta t$ for the barostat. The simulation parameters are summarised in Table \ref{tab:LJnptsim}. \\

\begin{table}[h!]
    \centering
    \begin{tabular}{c|c|c|c|c|c|c|c}
    $\Delta t$ & stride & MD data size & $\Delta t_T$ & $\Delta t_P$\\
    \hline 
    0.001 & 100 & 9.900 & $100 \Delta t$ & $1000 \Delta t$ 
    \end{tabular}
    \caption{MD simulation parameters for the LJ system with periodic boundary conditions in the $NPT$ ensemble, $\Delta t$ is the time step size, stride is the number of MD steps between saved configurations,  MD data size is the number of generated samples excluding discarded samples, and $\Delta t_T$ and $\Delta t_P$ are the time constants of the thermostat and barostat, respectively. }
    \label{tab:LJnptsim}
\end{table}

\subsection{Hemmer-Stell-like system in the $NPT$ ensemble}
\label{sec:apphsl}

In this section, we present details on the generation and training of the HSL system with periodic boundary conditions in the $NPT$ ensemble. We provide the parameters for the HSL model and the MD simulation, as well as the architecture and training schedule of the BG.

\subsubsection{Boltzmann generator protocol}

The parameters for the HSL potential, cut-off distance, regularization and the center-of-mass restraint are the same as for the LJ system and specified in Table \ref{tab:LJnpt_sys}. The parameters for the exponential well term as specified in Table \ref{tab:HSLnpt_sys}.\color{black}

\begin{table}[h!]
    \centering
    \begin{tabular}{c|c|c}
    $A$ & $w$ & $r_0$  \\
    \hline 
    1.5$\epsilon$ & 41.22$\sigma^{-2}$ & 1.44$\sigma$
    \end{tabular}
    \caption{Model parameters for the $NPT$ HSL system, where $A$ and $w$ are parameters of the exponential well centered at $r_0$. All values are in reduced units $\sigma$ and $\epsilon$, which are the particle diameter and the interaction strength respectively.  Values are taken from Ref. \cite{quigley_progression_2005}. The LJ paramters remain the same as in Table \ref{tab:LJnpt_BG}.}
    \label{tab:HSLnpt_sys}
\end{table}

As for the LJ system, we train an $NPT$ BG on KL loss only. The architecture of the BG is discussed in Section \ref{sec:bgs} and \ref{sec:npt_bg}. The hyperparameters are the same as in  Table \ref{tab:LJnpt_BG}. The BG for $P\sigma^2/\epsilon=5.0$ is trained with the schedule described in Table \ref{tab:HSLnpt_schedule}. For the BG at $P\sigma^2/\epsilon=2.0$, iter is changed to 6000 for all epochs.

\begin{table}[h!]
    \centering
    \begin{tabular}{c||c|c|c}
    epoch & 1 & 2 & 3  \\
    \hline 
    \hline iter & 4000 & 4000 & 4000\\
    \hline batch & 256 & 256 & 256  \\
    \hline lr & $10^{-3}$ & $10^{-4}$ & $10^{-5}$\\
    \hline $w_\text{M L}$ & 0 & 0 & 0 \\
    \hline $w_\text{K L}$ & 1 & 1 & 1
    \end{tabular}
    \caption{Training schedule for the $NPT$ BG for the HSL system at $P\sigma^2/\epsilon=5.0 $, iter is the number of iterations used for training, batch is the batch size, lr is the learning rate used in the Adam optimiser and $w_\text{ML}$, and $w_\text{KL}$ are the weights for the ML and KL loss respectively.  The schedule consists of 3 epochs. Each epoch has different hyperparameters.  For the BG at $P\sigma^2/\epsilon=2.0$, iter is changed to 6000 for all epochs.} 
    \label{tab:HSLnpt_schedule}
\end{table}

\subsubsection{Molecular dynamics protocol}

We generate MD data by running simulations using LAMMPS.  \cite{thompson_lammps_2022} The MD data is used only for comparison against the BG data. We initialise the coordinates in the same lattice as the LJ system. We simulate for $10^6$ MD steps with time step $\Delta t = 0.001$, saving a configuration every $100$ steps. Furthermore, we remove the first $100$ configurations. Therefore, the MD data consists of $9900$ samples. The thermostat \cite{martyna_nosehoover_1992} and barostat \cite{parrinello_polymorphic_1981} settings are the same as for the LJ case. We run two simulations, one at $P\sigma^2/\epsilon=2.0$ and another at $P\sigma^2/\epsilon=5.0$

\newpage

\bibliographystyle{plain}
\bibliography{main.bib}

\begin{thebibliography}{10}

\bibitem{ahmad_free_2022}
Rasool Ahmad and Wei Cai.
\newblock Free energy calculation of crystalline solids using normalizing
  flows.
\newblock {\em Modelling and Simulation in Materials Science and Engineering},
  30(6):065007, July 2022.
\newblock Publisher: IOP Publishing.

\bibitem{alder1959studies}
Berni~J Alder and Thomas~Everett Wainwright.
\newblock {Studies in molecular dynamics. I. General method}.
\newblock 31(2):459--466, 1959.

\bibitem{barker_phase_1981}
J.~A. Barker, D.~Henderson, and F.~F. Abraham.
\newblock Phase diagram of the two-dimensional {Lennard}-{Jones} system;
  {Evidence} for first-order transitions.
\newblock {\em Physica A: Statistical Mechanics and its Applications},
  106(1):226--238, March 1981.

\bibitem{boltzmann1898vorlesungen}
Ludwig Boltzmann.
\newblock {\em Vorlesungen {\"u}ber Gastheorie: 2. Teil}.
\newblock J.B. Barth, Leipzig, Germany, 1898.

\bibitem{dibak2022temperature}
Manuel Dibak, Leon Klein, Andreas Kr{\"a}mer, and Frank No{\'e}.
\newblock Temperature steerable flows and boltzmann generators.
\newblock {\em Physical Review Research}, 4(4):L042005, 2022.

\bibitem{dinh_density_2017}
Laurent Dinh, Jascha Sohl-Dickstein, and Samy Bengio.
\newblock Density estimation using {Real} {NVP}, February 2017.
\newblock arXiv:1605.08803 [cs, stat].

\bibitem{falkner2022conditioning}
Sebastian Falkner, Alessandro Coretti, Salvatore Romano, Phillip Geissler, and
  Christoph Dellago.
\newblock Conditioning normalizing flows for rare event sampling.
\newblock {\em arXiv preprint arXiv:2207.14530}, 2022.

\bibitem{frenkel_understanding_2002}
Daan Frenkel and Berend Smit.
\newblock {\em Understanding molecular simulation: from algorithms to
  applications}.
\newblock Number~1 in Computational science series. Academic Press, San Diego,
  2nd ed edition, 2002.

\bibitem{gabrie_adaptive_2022}
Marylou Gabrié, Grant~M. Rotskoff, and Eric Vanden-Eijnden.
\newblock Adaptive {Monte} {Carlo} augmented with normalizing flows.
\newblock {\em Proceedings of the National Academy of Sciences},
  119(10):e2109420119, March 2022.
\newblock Publisher: Proceedings of the National Academy of Sciences.

\bibitem{henin2022enhanced}
J{\'e}r{\^o}me H{\'e}nin, Tony Leli{\`e}vre, Michael~R Shirts, Omar Valsson,
  and Lucie Delemotte.
\newblock Enhanced sampling methods for molecular dynamics simulations.
\newblock {\em arXiv preprint arXiv:2202.04164}, 2022.

\bibitem{hsu_boltzmann_2022}
Wei-Tse Hsu and Theodore Fobe.
\newblock Boltzmann {Generators}, September 2022.
\newblock original-date: 2020-02-28T04:28:14Z.

\bibitem{jing2022torsional}
Bowen Jing, Gabriele Corso, Jeffrey Chang, Regina Barzilay, and Tommi Jaakkola.
\newblock Torsional diffusion for molecular conformer generation.
\newblock {\em arXiv preprint arXiv:2206.01729}, 2022.

\bibitem{kobyzev_normalizing_2021}
Ivan Kobyzev, Simon J.~D. Prince, and Marcus~A. Brubaker.
\newblock Normalizing {Flows}: {An} {Introduction} and {Review} of {Current}
  {Methods}.
\newblock {\em IEEE Transactions on Pattern Analysis and Machine Intelligence},
  43(11):3964--3979, November 2021.
\newblock arXiv: 1908.09257.

\bibitem{kramer_bgflow_2022}
Andreas Kramer, Jonas Kohler, Leon Klein, Frank Noe, Michele Invernizzi, and
  Dibak Manuel.
\newblock bgflow, September 2022.
\newblock original-date: 2021-04-15T15:37:40Z.

\bibitem{kohler_equivariant_2020}
Jonas Köhler, Leon Klein, and Frank Noé.
\newblock Equivariant {Flows}: {Exact} {Likelihood} {Generative} {Learning} for
  {Symmetric} {Densities}.
\newblock {\em arXiv:2006.02425 [physics, stat]}, October 2020.
\newblock arXiv: 2006.02425 version: 2.

\bibitem{laio_escaping_2002}
Alessandro Laio and Michele Parrinello.
\newblock Escaping free-energy minima.
\newblock {\em Proceedings of the National Academy of Sciences},
  99(20):12562--12566, October 2002.
\newblock Publisher: Proceedings of the National Academy of Sciences.

\bibitem{martyna_nosehoover_1992}
Glenn~J. Martyna, Michael~L. Klein, and Mark Tuckerman.
\newblock Nosé–{Hoover} chains: {The} canonical ensemble via continuous
  dynamics.
\newblock {\em The Journal of Chemical Physics}, 97(4):2635--2643, August 1992.
\newblock Publisher: American Institute of Physics.

\bibitem{metropolis1953equation}
Nicholas Metropolis, Arianna~W Rosenbluth, Marshall~N Rosenbluth, Augusta~H
  Teller, and Edward Teller.
\newblock Equation of state calculations by fast computing machines.
\newblock 21(6):1087--1092, 1953.

\bibitem{midgley2022flow}
Laurence~Illing Midgley, Vincent Stimper, Gregor~NC Simm, Bernhard
  Sch{\"o}lkopf, and Jos{\'e}~Miguel Hern{\'a}ndez-Lobato.
\newblock Flow annealed importance sampling bootstrap.
\newblock {\em arXiv preprint arXiv:2208.01893}, 2022.

\bibitem{noe_boltzmann_2019}
Frank Noé, Simon Olsson, Jonas Köhler, and Hao Wu.
\newblock Boltzmann generators: {Sampling} equilibrium states of many-body
  systems with deep learning.
\newblock {\em Science}, 365(6457):eaaw1147, September 2019.

\bibitem{papamakarios_normalizing_2021}
George Papamakarios, Eric Nalisnick, Danilo~Jimenez Rezende, Shakir Mohamed,
  and Balaji Lakshminarayanan.
\newblock Normalizing {Flows} for {Probabilistic} {Modeling} and {Inference}.
\newblock {\em arXiv:1912.02762 [cs, stat]}, April 2021.
\newblock arXiv: 1912.02762.

\bibitem{parrinello_polymorphic_1981}
M.~Parrinello and A.~Rahman.
\newblock Polymorphic transitions in single crystals: {A} new molecular
  dynamics method.
\newblock {\em Journal of Applied Physics}, 52(12):7182--7190, December 1981.
\newblock Publisher: American Institute of Physics.

\bibitem{quigley_progression_2005}
D.~Quigley and M.~I.~J. Probert.
\newblock Progression of phase behavior for a sequence of model core-softened
  potentials.
\newblock {\em Physical Review E}, 72(6):061202, December 2005.
\newblock Publisher: American Physical Society.

\bibitem{satorras_en_2021}
Victor~Garcia Satorras, Emiel Hoogeboom, Fabian~B. Fuchs, Ingmar Posner, and
  Max Welling.
\newblock E(n) {Equivariant} {Normalizing} {Flows}.
\newblock {\em arXiv:2105.09016 [physics, stat]}, June 2021.
\newblock arXiv: 2105.09016.

\bibitem{thompson_lammps_2022}
Aidan~P. Thompson, H.~Metin Aktulga, Richard Berger, Dan~S. Bolintineanu,
  W.~Michael Brown, Paul~S. Crozier, Pieter J. in~'t Veld, Axel Kohlmeyer,
  Stan~G. Moore, Trung~Dac Nguyen, Ray Shan, Mark~J. Stevens, Julien Tranchida,
  Christian Trott, and Steven~J. Plimpton.
\newblock {LAMMPS} - a flexible simulation tool for particle-based materials
  modeling at the atomic, meso, and continuum scales.
\newblock {\em Computer Physics Communications}, 271:108171, 2022.

\bibitem{torrie_nonphysical_1977}
G.~M. Torrie and J.~P. Valleau.
\newblock Nonphysical sampling distributions in {Monte} {Carlo} free-energy
  estimation: {Umbrella} sampling.
\newblock {\em Journal of Computational Physics}, 23(2):187--199, February
  1977.

\bibitem{winter_unsupervised_2022}
Robin Winter, Marco Bertolini, Tuan Le, Frank Noé, and Djork-Arné Clevert.
\newblock Unsupervised {Learning} of {Group} {Invariant} and {Equivariant}
  {Representations}, September 2022.
\newblock arXiv:2202.07559 [cs].

\bibitem{wirnsberger_targeted_2020}
Peter Wirnsberger, Andrew~J. Ballard, George Papamakarios, Stuart Abercrombie,
  Sébastien Racanière, Alexander Pritzel, Danilo Jimenez~Rezende, and Charles
  Blundell.
\newblock Targeted free energy estimation via learned mappings.
\newblock {\em The Journal of Chemical Physics}, 153(14):144112, October 2020.
\newblock Publisher: American Institute of Physics.

\bibitem{wirnsberger_normalizing_2021}
Peter Wirnsberger, George Papamakarios, Borja Ibarz, Sébastien Racanière,
  Andrew~J. Ballard, Alexander Pritzel, and Charles Blundell.
\newblock Normalizing flows for atomic solids.
\newblock {\em arXiv:2111.08696 [cond-mat, physics:physics, stat]}, November
  2021.
\newblock arXiv: 2111.08696.

\bibitem{zhu2002using}
Zhongwei Zhu, Mark~E Tuckerman, Shane~O Samuelson, and Glenn~J Martyna.
\newblock Using novel variable transformations to enhance conformational
  sampling in molecular dynamics.
\newblock {\em Physical review letters}, 88(10):100201, 2002.

\bibitem{zwanzig1954high}
Robert~W Zwanzig.
\newblock High-temperature equation of state by a perturbation method. i.
  nonpolar gases.
\newblock {\em The Journal of Chemical Physics}, 22(8):1420--1426, 1954.

\end{thebibliography}


\end{document}